\documentclass[aps,twocolumn,showpacs,amsmath,amssymb,prc]{revtex4-2}
\usepackage{graphicx,here}
\usepackage{multirow}
\usepackage{tikz}
\usepackage{epstopdf}
\usepackage[percent]{overpic}
\usepackage{scrextend}
\usepackage{xcolor,xspace}
\usepackage[ulem=normalem]{changes}
\usepackage{hyperref}
\hypersetup{colorlinks=True,urlcolor=blue,linkcolor=blue,citecolor=blue,filecolor=black}

\colorlet{Changes@Color}{red}
\usepackage{dcolumn,color,footnote,bm,braket}
\usepackage{url,longtable,tabularx}
\usepackage{threeparttable}

\usepackage{fancybox}
\usetikzlibrary{matrix}

\newcommand\+{\dagger}

\begin{document}

\title{Role of octupole shape degree of freedom in neutron-rich odd-mass xenon isotopes}

\author{K. Nomura}
\email{nomura@sci.hokudai.ac.jp}
\affiliation{Department of Physics, 
Hokkaido University, Sapporo 060-0810, Japan}
\affiliation{Nuclear Reaction Data Center, 
Hokkaido University, Sapporo 060-0810, Japan}

\date{\today}

\begin{abstract}
Influences of the octupole shape degree of freedom on 
low-energy spectra of neutron-rich odd-mass xenon isotopes 
are studied within the interacting boson-fermion model 
that is based on the nuclear density functional theory. 
The interacting-boson Hamiltonian describing low-energy 
quadrupole and octupole collective states of the 
even-even nuclei $^{140,142,144}$Xe, single-particle 
energies, and occupation probabilities 
for an unpaired neutron 
in the odd-mass nuclei $^{141,143,145}$Xe, 
are determined based on 
the axially symmetric quadrupole-octupole 
deformation-constrained 
self-consistent mean-field calculations 
with a choice of the energy density functional 
and pairing interaction. 
Strength parameters of the boson-fermion interactions 
are empirically determined to reproduce a 
few low-lying levels of each odd-mass nucleus. 
The mean-field calculation predicts for $^{142}$Xe 
a potential energy surface that is notably soft in the octupole 
deformation with a non-zero octupole global minimum. 
The octupole correlations are shown to 
be relevant in positive-parity excited states of $^{143,145}$Xe. 
\end{abstract}

\maketitle

\section{Introduction}

Search for the reflection asymmetric, octupole deformation 
in nuclei has been a recurrent theme of interest 
that is under active investigations for several decades. 
The octupole correlations are considered to be 
enhanced in several specific mass regions 
corresponding to the neutron $N$ and/or 
proton $Z$ numbers equal to 34, 56, 88, 134, $\ldots$, 
in which coupling between the normal and unique-parity 
single-particle orbitals differing 
in the total and orbital angular momenta 
by $3\hbar$ is possible \cite{butler1996}. 
Most of those nuclei 
with the aforementioned nucleon numbers 
are short lived, but have recently 
become accessible by experiments using radioactive beams. 
In addition, since observation of a large 
electric dipole moment in the octupole-deformed nucleus 
would indicate violation of time-reversal (T) 
or charge-parity (CP) symmetry, it is also relevant 
to a possible extension of the standard model 
of particle physics \cite{engel2013}. 
Evidence for a static octupole 
deformation was found in light actinide region, 
i.e., $^{220}$Rn and $^{224}$Ra, 
by the experiment at ISOLDE, CERN \cite{gaffney2013}. 
The experiments carried out 
at Argonne National Laboratory 
found another candidates for the 
octupole deformation in the 
neutron-rich region, i.e., 
$^{144}$Ba \cite{bucher2016} 
and $^{146}$Ba \cite{bucher2017}.

The present study is focused on the octupole correlation 
effects in the neutron-rich even- and odd-mass Xe isotopes 
in the ``north east'' of the 
doubly magic nucleus $^{132}$Sn. 
Since in this mass region octupole collectivity 
is supposed to be most pronounced 
in the neighborhood of the $^{144}$Ba nucleus 
(with $N=88$ and $Z=56$), 
it is of interest to see if the octupole degree 
of freedom still plays a role in the Xe isotopes 
with $N\approx 88$. 
Spectroscopic data for the 
neutron-rich Xe isotopes are also available. 
For example, 
in the $\beta$-decay study performed as part 
of the EURICA project in RIKEN, level structure 
of the neutron-rich $^{140}$Xe nucleus was 
determined \cite{yagi2022}, and this 
study has been extended further to those 
of the odd-mass nuclei 
$^{141,143}$Xe \cite{nor2024phd,nor2024}. 
The fact that new data both 
on the even-even and odd-even 
neutron-rich Xe isotopes have become available 
thus necessitates theoretical investigations in a timely manner, 
that allow for a systematic as well as quantitative 
prediction on the details of their low-lying structures.

The theoretical method employed in the present 
study is the interacting boson model (IBM) that 
is based on the nuclear density functional theory. 
The starting point is the 
constrained self-consistent mean-field (SCMF) 
calculations with a choice of the 
energy density functional (EDF) 
and pairing interaction, providing the 
potential energy surface (PES) as a function 
of the axially symmetric 
quadrupole and octupole deformations. 
The quadrupole-octupole SCMF PES is 
then mapped onto the expectation 
value of the Hamiltonian of the IBM, 
that consists of 
monopole $s$ (with spin and parity $J^\pi=0^+$), 
quadrupole $d$ ($J^\pi=2^+$), 
and octupole $f$ ($J^\pi=3^-$) bosons. 
The mapping procedure specifies the $sdf$-IBM 
Hamiltonian describing low-lying positive- 
and negative-parity states of even-even nuclei. 
Odd-mass nuclei are dealt with by means 
of the particle-boson coupling in the 
interacting bosom-fermion model (IBFM) \cite{IBFM}, 
with the even-even $sdf$-IBM core Hamiltonian 
determined by the mapping procedure. 
The same SCMF calculations provide microscopic 
input to determine most parts of 
the $sdf$-IBFM Hamiltonian. 
Strength parameters for the boson-fermion 
interactions are, however, obtained empirically 
to reproduce low-energy 
spectra of each odd-mass nucleus.

The mapped $sdf$-IBM has been extensively used 
to study low-energy quadrupole and octupole 
collective states for those even-even nuclei 
in light actinide 
\cite{nomura2013oct,nomura2014,nomura2023oct}, 
rare-earth \cite{nomura2014}, 
lanthanide \cite{nomura2014,nomura2021oct-ba}, 
neutron-rich $A \approx 100$ \cite{nomura2022-oct56}, 
and neutron-deficient 
$N \approx Z \approx 34$ \cite{nomura2022octcm} regions. 
In Ref.~\cite{nomura2018oct}, in particular, 
the $sdf$-IBFM framework has been developed 
and applied to the 
neutron-rich odd-mass $^{143,145,147}$Ba nuclei 
with the microscopic input obtained from the 
relativistic EDF. 
The $sdf$-IBFM 
produced for $^{143,145,147}$Ba low-energy band structures 
in agreement with data \cite{rzacaurban2012,data}, 
and suggested some excited bands 
that can be interpreted as octupole bands 
connected by the 
$E3$ transitions to ground-state bands. 
Earlier experiments on the $^{141,143}$Xe nuclei
\cite{urban2000-141Xe,huang2017-141Xe,rzacaurban2011} 
suggested more or less 
similar low-lying band structures to those observed 
for the neighboring $^{143,145}$Ba nuclei. 
It is thus feasible to extend the $sdf$-IBFM 
calculation to the Xe isotopes, since similar 
model parameters to those used for the odd-mass Ba 
could be employed.

In Sec.~\ref{sec:model} the theoretical procedure is 
described. Results of the mean-field calculations 
and spectroscopic calculations within the $sdf$-IBM 
for the even-even Xe isotopes are discussed 
in Sec.~\ref{sec:even}. 
In Sec.~\ref{sec:odd} calculated energy levels 
and transition properties for the odd-mass Xe nuclei within 
the $sdf$-IBFM are reported. 
Summary of the main results and conclusions are 
given in Sec.~\ref{sec:summary}.

\section{Theoretical framework\label{sec:model}}

The constrained SCMF calculations are 
carried out within the relativistic 
Hartree-Bogoliubov (RHB) model \cite{vretenar2005,niksic2011} 
using the density-dependent point-coupling (DD-PC1) \cite{DDPC1} 
functional for particle-hole channel and 
a separable pairing force of finite range 
\cite{tian2009} in the particle-particle channel. 
The constraints are those on the axially symmetric mass 
quadrupole $Q_{20}$ and octupole $Q_{30}$ moments, 
which are related to the dimensionless 
shape variables $\beta_2$ and $\beta_3$, respectively:
\begin{eqnarray}
 \beta_\lambda = \frac{4\pi}{3AR^{\lambda}}Q_{\lambda 0}
\end{eqnarray}
with $\lambda=2$ or 3, and $R=1.2A^{1/3}$ fm.  
The RHB-SCMF calculations produce 
for the even-even 
$^{140,142,144}$Xe nuclei the $(\beta_2,\beta_3)$-PESs, 
which are used to determine the $sdf$-IBM Hamiltonian 
by the procedure described below.

The even-even core nuclei are here described 
in terms of the interacting $s$, $d$, 
and $f$ bosons \cite{engel1985,barfield1988}, 
which are microscopically interpreted as 
the monopole, quadrupole, and octupole 
collective pairs of valence nucleons, respectively, 
as in the $sd$-IBM framework \cite{IBM,OAIT,OAI}. 
The odd-mass nucleus is described as a coupled 
system of the even-even IBM core and an unpaired nucleon. 
The $sdf$-IBFM Hamiltonian is given in general as
\begin{eqnarray}
\label{eq:ham}
 \hat H_{\rm IBFM} = \hat H_{\rm B} + \hat H_{\rm F} + \hat V_{\rm BF} \; ,
\end{eqnarray}
where the first, second, and third terms on 
the right-hand side represent the $sdf$-IBM Hamiltonian, 
single-fermion Hamiltonian, and 
boson-fermion interaction, respectively. 
Here $\hat H_{\rm B}$ takes the form
\begin{align}
\label{eq:bh}
 \hat H_{\rm B} = 
\epsilon_{d} \hat n_{d} +\epsilon_{f} \hat n_{f} 
  + \kappa_{2} \hat Q \cdot \hat Q
  + \kappa_{3} \hat O \cdot \hat O 
  + \kappa' \hat L \cdot \hat L \; .
\end{align}
In the first (second) term, $\hat n_d=d^\+\cdot\tilde d$ 
($\hat n_f = f^\+ \cdot \tilde f$), 
with $\epsilon_{d}$ ($\epsilon_{f}$) representing 
the single $d$ ($f$) boson energy relative to the 
$s$-boson one. 
Note $\tilde d_{\mu} = (-1)^{\mu} d_{-\mu}$ 
and $\tilde f_{\mu} = (-1)^{3+\mu} f_{-\mu}$. 
The third and fourth terms in (\ref{eq:bh}) stand for 
quadrupole-quadrupole and octupole-octupole 
interactions, respectively. 
The quadrupole $\hat Q$ and octupole $\hat O$ 
operators read
\begin{align}
\label{eq:quad}
& \hat Q = s^\+ \tilde d + d^\+ s 
+ \chi (d^\+ \times \tilde d)^{(2)}
+ \chi' (f^\+ \times \tilde f)^{(2)} \; , 
\\
\label{eq:oo}
& \hat O = s^\+\tilde f + f^\+ s
+ \chi''(d^\+\times\tilde f +f^\+\times\tilde d)^{(3)} \; ,
\end{align}
with $\chi$, $\chi'$, and $\chi''$ being 
dimensionless parameters. 
The last term in Eq.~(\ref{eq:bh}), 
$\hat L\cdot \hat L$, is introduced 
to better describe moments of inertia 
of yrast bands, with $\hat L$ being the boson 
angular momentum operator 
\begin{align}
 \hat L=\sqrt{10}(d^\+\times\tilde d)^{(1)}-\sqrt{28}(f^\+\times\tilde f)^{(1)} \; .
\end{align} 

The form of the $sdf$-IBM Hamiltonian 
in Eq.~(\ref{eq:bh}) is slightly different 
from that employed in the previous study 
on odd-mass Ba \cite{nomura2018oct}: 
in Ref.~\cite{nomura2018oct} 
the octupole-octupole term was taken to be of the form, 
$:\hat V_3^\+ \cdot \hat V_3:$, where 
$\hat V_3^\+ = s^\+\tilde f + \chi''(d^\+\times\tilde f)^{(3)}$
and the notation $:():$ represents normal ordering, 
and the angular momentum operator in the 
$\hat L \cdot \hat L$ term 
consisted only of the $d$-boson term, i.e., 
$\hat L=\sqrt{10}(d^\+\times\tilde d)^{(1)}$.

The boson analog of the $(\beta_2,\beta_3)$-PES 
is given by the energy expectation value 
$E_{\rm IBM}(\beta_2,\beta_3)=\bra{\phi(\beta_2,\beta_3)}\hat H_{\rm B}\ket{\phi(\beta_2,\beta_3)}$. 
Here $\ket{\phi(\beta_2,\beta_3)}$ represents 
the coherent state \cite{ginocchio1980} 
of $s$, $d$, and $f$ bosons, given by
\begin{eqnarray}
 \ket{\phi(\beta_2,\beta_3)} = (s^\+ + \bar\beta_2 d_0^\+ + \bar\beta_3 f_0^\+)^{n}\ket{0} \; ,
\end{eqnarray}
up to a normalization factor, 
where $n$ is the number of bosons, i.e., valence nucleon pairs, 
$\bar\beta_\lambda$ is a boson analog of the 
quadrupole ($\lambda=2$) or octupole ($\lambda=3$) 
deformation variable in the geometrical model \cite{BM}, 
and $\ket{0}$ represents the inert core, or 
the doubly-magic nucleus $^{132}$Sn. 
It is assumed that bosonic deformation is proportional 
to the geometrical one, that is, 
$\bar\beta_\lambda=C_\lambda \beta_\lambda$, 
where $C_\lambda$ stands for a constant of 
proportionality. 

The $sdf$-IBM Hamiltonian (\ref{eq:bh}) 
is determined for 
each nucleus by mapping the RHB-SCMF PES, 
$E_{\rm SCMF}(\beta_2,\beta_3)$, 
onto the bosonic counterpart, $E_{\rm IBM}(\beta_2,\beta_3)$, 
so that both energy surfaces become similar 
in topology in the neighborhood of the 
global minimum \cite{nomura2008,nomura2010,nomura2014}, 
$E_{\rm SCMF}(\beta_2,\beta_3) \approx E_{\rm IBM}(\beta_2,\beta_3)$. 
The mapping procedure determines the strength 
parameters for $\hat H_{\rm B}$ together with 
the constants $C_2$ and $C_3$. 
However, since the $\hat L\cdot \hat L$ term does not 
make a contribution to the PES, its strength parameter 
$\kappa'$ has to be determined in a 
separately way, so that the moment of inertia 
calculated in the boson intrinsic state 
\cite{schaaser1986} at the 
equilibrium minimum should 
be equal to the corresponding 
cranking moment of inertia 
computed with the RHB-SCMF method using 
the Inglis-Belyaev (IB) formula 
\cite{inglis1956,belyaev1961}. 
The IB moment of inertia is here 
increased by 30 \% in order to account for 
the fact the IB formula significantly 
underestimates the empirical moments of inertia. 
See Ref.~\cite{nomura2011rot} for further details. 
Table~\ref{tab:paraB} lists the parameters for $\hat H_{\rm B}$ 
adopted in the present study.

\begin{table}[hb!]
\caption{\label{tab:paraB} The parameters of the $sdf$-IBM
 Hamiltonian. $\epsilon_d$, $\epsilon_f$, $\kappa_2$,
 $\kappa'$ and $\kappa_3$ are in units of MeV, 
and the others are dimensionless.}
\begin{center}
 \begin{ruledtabular}
\begin{tabular}{ccccccccc}
\textrm{} &
\textrm{$\epsilon_d$} &
\textrm{$\epsilon_f$}&
\textrm{$\kappa_2$} &
\textrm{$\chi$}&
\textrm{$\chi'$}&
\textrm{$\kappa'$} &
\textrm{$\kappa_3$}&
\textrm{$\chi''$}\\
\hline
$^{140}$Xe & 0.881 & 0.852 & $-0.098$ & $-1.2$ & $-1.7$ & $-0.023$ & 0.030 & $-0.8$ \\
$^{142}$Xe & 0.644 & 0.710 & $-0.098$ & $-1.3$ & $-2.2$ & $-0.016$ & 0.048 & $-1.6$ \\
$^{144}$Xe & 0.661 & 0.951 & $-0.098$ & $-1.3$ & $-2.0$ & $-0.018$ & 0.048 & $-1.6$ \\
\end{tabular}
 \end{ruledtabular}
\end{center}
\end{table}

The single-fermion Hamiltonian in Eq.~(\ref{eq:ham}) 
for the odd neutron reads,  
$\hat H_{\rm F}=\sum_j \epsilon_j (a_j^\+ \times \tilde a_j)^{(0)}$, 
where $\epsilon_j$ is the single-particle energy, 
$a_j^\+$ is creation operator for a particle in 
orbital $j$, and $\tilde a_j \equiv (-1)^{j-m} a_{j-m}$ 
denotes annihilation operator. 
The fermion valence space 
for the considered odd-mass nuclei $^{141,143,145}$Xe 
comprises all single-particle
levels in the neutron $N=82-126$ major oscillator shell, 
that is, $3p_{1/2}$,
$3p_{3/2}$, $2f_{5/2}$, $2f_{7/2}$, $1h_{9/2}$ and $1i_{13/2}$.

The boson-fermion interaction in Eq.~(\ref{eq:ham}), 
$\hat V_{\rm BF}$, consists of the terms that 
represent the coupling of the odd neutron to the 
$sd$-boson space $\hat V_{\rm BF}^{sd}$,
to the $f$ boson space $\hat V_{\rm BF}^{f}$,
and to the combined $sdf$-boson space $\hat V_{\rm BF}^{sdf}$:
\begin{eqnarray}
\label{eq:bf}
 \hat V_{\rm BF}=\hat V_{\rm BF}^{sd} + \hat V_{\rm BF}^{f} + \hat V_{\rm BF}^{sdf}. 
\end{eqnarray}
The first term in Eq.~(\ref{eq:bf}) reads: 
\begin{eqnarray}
 \label{eq:hbf-sd}
 \hat V_{\rm BF}^{sd}
=&&\sum_{j_aj_b}\Gamma_{j_aj_b}^{sd}\hat
  Q_{sd}\cdot (a^{\dagger}_{j_a}\times\tilde a_{j_b})^{(2)}
\nonumber \\
&&+\sum_{j_aj_bj_c}\Lambda_{j_aj_bj_c}^{dd}
:\left[(a_{j_a}^\+\times\tilde d)^{(j_c)} 
\times 
(d^\+\times\tilde a_{j_b})^{(j_c)}\right]^{(0)}:
\nonumber \\
&&
+\sum_{j_a} A_{j_a}^d
(a^{\+}_{j_a}\times\tilde a_{j_a})^{(0)}\hat n_d, 
\end{eqnarray}
where the first, second and third 
terms denote quadrupole 
dynamical, exchange and monopole terms, 
respectively \cite{scholten1985,IBFM}. 
$\hat Q_{sd}$ is the $sd$ part of the quadrupole operator 
in Eq.~(\ref{eq:quad}). 
In this section, 
single-particle orbitals are denoted by 
$j_a, j_b, j_c, \ldots$, while primed ones, such as $j_a', j_b', j_c' \ldots$,
stand for those with opposite parity, unless otherwise specified. 
In a similar fashion to Eq.~(\ref{eq:hbf-sd})
the following form is considered 
for the $f$-boson part:
\begin{eqnarray}
 \hat V_{\rm BF}^f 
&&= \sum_{j_aj_b}\Gamma^{ff}_{j_aj_b}\hat
  Q_{ff}\cdot (a^{\dagger}_{j_a}\times\tilde a_{j_b})^{(2)}
\nonumber \\
&&+
\sum_{{j_a}{j_b}{j_c^{\prime}}}
\Lambda_{j_aj_bj_c^{\prime}}^{ff}
:\left[(a^\+_{j_a}\times\tilde f)^{(j_c^{\prime})}
\times
(f^\+\times \tilde a_{j_b})^{(j_c^{\prime})}\right]^{(0)}:
\nonumber \\
&&
+\sum_{{j_a}}A_{j_a}^{f}
(a^\+_{j_a}\times\tilde a_{j_a})^{(0)}\hat n_f \; ,
\end{eqnarray}
where $\hat Q_{ff}=\chi' (f^\+\times\tilde f)^{(2)}$ 
is identified as the fourth term of 
the quadrupole operator $\hat Q$ [Eq.~(\ref{eq:quad})]. 
Finally, $\hat V_{\rm BF}^{sdf}$ in Eq.~(\ref{eq:bf}) reads:  
\begin{eqnarray}
\label{eq:hbf-sdf}
\hat V_{\rm BF}^{sdf} 
&&
= 
\sum_{{j_a}{j_b^{\prime}}}\Gamma_{{j_a}{j_b^{\prime}}}^{sdf}
\hat O\cdot [a_{j_a}^{\dagger}\times\tilde a_{j_b^{\prime}}]^{(3)}
\nonumber \\
&&+
\sum_{{j_a}{j_b^{\prime}j_c}}
\Lambda_{j_aj_b^{\prime}j_c}^{df}
:\left[
[a_{j_a}^\+\times\tilde d]^{(j_c)}
\times
[f^\+\times \tilde a_{j_b^{\prime}}]^{(j_c)}
\right]^{(0)}:
\nonumber \\
&&
+ ({\rm H.c.}), 
\end{eqnarray}
where the first and second terms denote 
the dynamical octupole and exchange terms, 
respectively.

By using the generalized seniority scheme 
\cite{scholten1985}, 
the $j$-dependent coefficients of the three terms in 
$\hat V^{sd}_{\rm BF}$ in Eq.~(\ref{eq:hbf-sd}) 
are shown to have the forms
\begin{eqnarray}
\label{eq:a-sd}
 &&A_{j}^{d}=-A_0^{d}\sqrt{2j+1} \\
\label{eq:gam-sd}
 &&\Gamma_{j_aj_b}^{sd}
=\Gamma_0^{sd}\gamma_{j_aj_b}^{(2)} \\
\label{eq:lam-sd}
 &&\Lambda_{j_aj_bj_c}^{dd}
=-2\Lambda_0^{dd}\sqrt{\frac{5}{2j_c+1}}
\beta_{j_aj_c}^{(2)}\beta_{j_bj_c}^{(2)} \; ,
\end{eqnarray}
with $\Gamma_0^{sd}$, $\Lambda_0^{sd}$, $A_0^d$ 
denoting strength parameters. 
The above formulas have been extended to 
$sdf$-boson systems in Ref.~\cite{nomura2018oct}, 
and similar expressions 
are obtained for the coefficients 
in $\hat V_{\rm BF}^f$ and $\hat V_{\rm BF}^{sdf}$. 
For the $f$-boson part:
\begin{eqnarray}
\label{eq:a-ff}
&&A_{j}^{f}=-A_0^{f}\sqrt{2j+1} \\
\label{eq:gam-ff}
&&\Gamma_{j_aj_b}^{ff}
=\Gamma_0^{ff}\gamma_{j_aj_b}^{(2)} \\
\label{eq:lam-ff}
&& \Lambda_{j_aj_bj_c'}^{ff}
=-2\Lambda_0^{ff}\sqrt{\frac{7}{2j_c^{\prime}+1}}
\beta_{j_aj_c'}^{(3)}\beta_{j_bj_c'}^{(3)}
\end{eqnarray}
and for the $sdf$-boson terms:
\begin{eqnarray}
\label{eq:gam-df}
 &&\Gamma_{j_aj_b^{\prime}}^{sdf}
=\Gamma_0^{sdf}\gamma_{j_aj_b'}^{(3)} \\
\label{eq:lam-df}
 &&\Lambda_{j_aj_b^{\prime}j_c}^{df}
=-2\Lambda_0^{df}\sqrt{\frac{7}{2j_c+1}}
\beta_{j_aj_c}^{(2)}\beta_{j_b'j_c}^{(3)} \; . 
\end{eqnarray}
In the expressions in 
Eqs.~(\ref{eq:a-sd})--(\ref{eq:lam-df}) 
$\gamma^{(\lambda)}_{ij}=(u_iu_{j}-v_iv_{j})q_{ij}^{(\lambda)}$ 
and
$\beta_{ij}^{(\lambda)}=(u_iv_{j}+u_{j}v_i)q_{ij}^{(\lambda)}$,
where 
$v_j$ ($u_j$) is an occupation (unoccupation) 
amplitude, and $q_{ij}^{(\lambda)}$ represents the matrix
element of fermion quadrupole ($\lambda=2$) 
or octupole ($\lambda=3$)
operator in the single-particle basis. 
Within this framework 
the single-particle energy $\epsilon_j$ should be 
replaced with the quasiparticle energy 
denoted by $\tilde\epsilon_j$. 
$\tilde\epsilon_j$ as well as 
occupation probabilities $v_j^2$ for the considered 
single-particle states are provided by the RHB-SCMF 
calculations constrained to zero deformation 
(see Refs.~\cite{nomura2016odd,nomura2018oct} for details). 
The adopted $\tilde\epsilon_j$ and 
$v_j^2$ are listed in Table~\ref{tab:spe}.

\begin{table}[hb!]
\caption{\label{tab:spe} 
Quasiparticle energies $\tilde\epsilon_j$ (in MeV) 
and occupation probabilities $v_j^2$ of 
the odd neutron 
for the orbitals $3p_{1/2}$, $3p_{3/2}$, 
$2f_{5/2}$, $2f_{7/2}$, $1h_{9/2}$, and 
$1i_{13/2}$ calculated for $^{141,143,145}$Xe 
with the RHB-SCMF method. 
}
\begin{center}
 \begin{ruledtabular}
\begin{tabular}{cccccccc}
& &
\textrm{$3p_{1/2}$} &
\textrm{$3p_{3/2}$} &
\textrm{$2f_{5/2}$} &
\textrm{$2f_{7/2}$} &
\textrm{$1h_{9/2}$} &
\textrm{$1i_{13/2}$} \\
\hline
\multirow{2}{*}{$^{141}$Xe} &
$\tilde\epsilon_j$ & 
3.163 & 2.688 & 2.840 & 1.285 & 1.392 & 0.981 \\
& $v_j^2$ & 
0.016 & 0.023 & 0.038 & 0.234 & 0.265 & 0.019 \\
[0.5em]
\multirow{2}{*}{$^{143}$Xe} &
$\tilde\epsilon_j$ & 
2.992 & 2.530 & 2.698 & 1.308 & 1.413 & 0.974 \\
& $v_j^2$ & 
0.024 & 0.034 & 0.053 & 0.315 & 0.376 & 0.027 \\
[0.5em]
\multirow{2}{*}{$^{145}$Xe} &
$\tilde\epsilon_j$ & 
2.814 & 2.366 & 2.546 & 1.328 & 1.454 & 0.966 \\
& $v_j^2$ & 
0.031 & 0.046 & 0.070 & 0.395 & 0.484 & 0.034 \\
\end{tabular}
 \end{ruledtabular}
\end{center}
\end{table}

There are 14 strength parameters for the boson-fermion
interaction $\hat V_{\rm BF}$ 
that have to be adjusted to the
spectroscopic data for the odd-mass Xe isotopes: 
six ($\Gamma_0^{sd}$,
$\Gamma_0^{ff}$, $\Lambda_0^{dd}$, 
$\Lambda_0^{ff}$, $A_0^d$ and $A_0^f$) for each of the 
normal-parity $pfh$ 
(i.e., $3p_{1/2,3/2}$, $2f_{5/2,7/2}$, 
$1h_{9/2}$) and the unique-parity
$1i_{13/2}$ single-particle configurations, 
and two additional parameters 
$\Gamma_0^{sdf}$ and $\Lambda_0^{df}$.
Their values are determined by taking the 
procedure described in the following.

First, the strength parameters 
$\Gamma_0^{sd}$, $\Lambda_0^{sd}$, and $A_0^d$ 
for the unique-parity ($1i_{13/2}$) and 
normal-parity ($pfh$) configurations 
are determined within the $sd$-IBM 
so as to reproduce, respectively, 
a few lowest-lying positive- and negative-parity 
levels for $^{141,143}$Xe. 
For $^{143}$Xe a $\Delta I=2$ band based 
on the $I={9/2}$ level without the parity being assigned 
is experimentally suggested \cite{rzacaurban2011}, 
and is here assumed to be a positive-parity band, 
based on the fact that a ${9/2}^+$ band at a similar 
excitation energy is observed in $^{145}$Ba \cite{rzacaurban2012}. 
The corresponding parameters 
$\Gamma_0^{sd}$, 
$\Lambda_0^{sd}$, and $A_0^d$ for the 
$1i_{13/2}$ single-particle configuration for $^{143}$Xe 
are determined so that a few low-lying levels 
of the parity-unassigned $I={9/2}$ band are 
reasonably reproduced. 
In addition,  
for the negative-parity states of both $^{141,143}$Xe, 
a special attention is paid to reproduce the order of 
the ${5/2}^-_1$, ${7/2}^-_1$, and ${9/2}^-_1$ 
levels. 
Note that any energy offset between 
the lowest negative- and positive-parity states, 
which is often introduced as a free parameter in the 
IBFM, is not introduced for all the odd-mass Xe 
nuclei studied in the present work.

The second step is to determine 
the parameters $\Gamma_0^{ff}$, 
$\Lambda_0^{ff}$, and $A_0^f$ within the $sdf$-IBFM. 
For the sake of simplicity, 
the same values 
as those used in Ref.~\cite{nomura2018oct} 
are adopted as an initial guess, and are  
readjusted so that the following conditions 
should be met for the odd-mass Xe nuclei: 
(i) the ground state spin should be 
${5/2}^-$ in the $sdf$-IBFM; 
(ii) for $^{141}$Xe, 
the ${13/2}^+_1$ and ${15/2}^+_1$ levels 
should be close to the measured ones \cite{data}; 
(iii) for $^{143}$Xe, the calculated ${13/2}^+_1$ level 
should be close to the measured ${13/2}$ state 
at 703 keV \cite{rzacaurban2011}, 
and the ${5/2}^+_1$ level appears at low energy, 
since in the $N=89$ isotone $^{145}$Ba 
several ${5/2}^+$ levels are observed at 
the excitation energies $E_x \approx 0.3-0.5$ MeV 
and a similar structure is expected to occur 
in $^{143}$Xe. 
The parameters 
$\Gamma_0^{sd}$, $\Lambda_0^{sd}$, and $A_0^d$ 
obtained in the previous step are 
also slightly modified to accommodate the 
above conditions.

Third, as was done in Ref.~\cite{nomura2018oct}, 
the parameters $\Gamma_0^{sdf}$ and $\Lambda_0^{df}$ 
are introduced only perturbatively as they 
turn out be of little importance for 
energy spectra. Here for simplicity a similar value to that in 
Ref.~\cite{nomura2018oct}, $\Gamma_0^{sdf}=0.15$ MeV, 
is employed, and $\Lambda_0^{df}=0$ MeV is assumed 
for all the odd-mass Xe nuclei under study. 
These values are taken to be the same for 
both parities in order to reduce the number of parameters.

Finally, the $sdf$-IBFM parameters for the $^{145}$Xe nucleus 
are determined so as to be more or less close to 
those determined for $^{141,143}$Xe, since experimental 
data are not available for $^{145}$Xe. 
Here it is assumed that 
the spin of the ground-state is ${5/2}^-$, 
and that the low-lying positive-parity levels 
exhibit a gradual decrease from $N=89$ to 91 
as in the case of the neighboring odd-mass Ba isotopes.

The adopted strength parameters 
for the $^{141,143,145}$Xe nuclei are listed 
in Table~\ref{tab:paraBF}. 
Most of these parameters exhibit only a
gradual variation with nucleon number. 
They are also more 
or less similar to those considered in 
the previous calculation 
for the odd-mass Ba nuclei \cite{nomura2018oct}. 
An only notable difference is 
that the present values of the parameters 
$\Gamma_0^{sd}=0.2$ MeV and $\Gamma_0^{ff}=0.2$ MeV 
for the unique-parity configuration 
are much smaller than those used in 
Ref.~\cite{nomura2018oct}, 
$\Gamma_0^{sd}=1.4$ MeV and $\Gamma_0^{ff}=1.2$ MeV.

The $sdf$-IBFM Hamiltonian with the 
parameters determined by the aforementioned procedure 
is numerically diagonalized \cite{arbmodel} 
within the model space consisting of 
$n$ $s$, $d$, and $f$ bosons and a 
single neutron in the full $pfh+i_{13/2}$ shells. 
In the present calculation, 
the maximum number of $f$ bosons, $n_f^{\rm max}$, 
is set equal to the total number of bosons, $n$, 
that is, no restriction 
is made of the $f$-boson number, 
whereas in the previous study of Ref.~\cite{nomura2018oct} 
it was limited to $n_f^{\rm max}=1$ in order to 
reduce computational time.

\begin{table}[htb!]
\caption{\label{tab:paraBF}
Strength parameters of the boson-fermion
 interaction $\hat V_{\rm BF}$ in Eq.~(\ref{eq:bf}) employed in the present
 calculation for the $^{141,143,145}$Xe nuclei (in MeV
 units). The numbers in the upper (lower) row for each nucleus correspond to 
 the negative-parity (positive-parity) single-particle configurations.}
\begin{center}
\begin{ruledtabular}
\begin{tabular}{ccccccccc}
\textrm{} &
\textrm{$\Gamma^{sd}_0$} &
\textrm{$\Gamma^{ff}_0$}&
\textrm{$\Lambda^{sd}_0$} &
\textrm{$\Lambda^{ff}_0$}&
\textrm{$A^{d}_0$}&
\textrm{$A^{f}_0$}&
\textrm{$\Gamma^{sdf}_0$}&
\textrm{$\Lambda^{df}_0$}\\
\hline
\multirow{2}{*}{$^{141}$Xe} 
& 0.20 & 0.20 & 0.0 & 0.0 & $-0.30$ & 0.0 & \multirow{2}{*}{0.15} & \multirow{2}{*}{0.0} \\
& 0.40 & 0.13 & 0.97 & 0.06 & $-0.60$ & $-0.15$ & & \\
[0.5em]
\multirow{2}{*}{$^{143}$Xe}
& 0.20 & 0.20 & 0.3 & 0.0 & $-1.5$ & 0.0 & \multirow{2}{*}{0.15} & \multirow{2}{*}{0.0} \\
& 0.40 & 0.60 & 0.3 & 0.0 & $-0.7$ & $-0.35$ & & \\
[0.5em]
\multirow{2}{*}{$^{145}$Xe} 
& 0.20 & 0.20 & 0.0 & 0.0 & $0.0$ & 0.0 & \multirow{2}{*}{0.15} & \multirow{2}{*}{0.0} \\
& 0.20 & 0.13 & 0.1 & 0.06 & $-0.4$ & $-0.15$ & & \\
\end{tabular}
\end{ruledtabular}
\end{center}
\end{table}

In the present work electric quadrupole 
($E2$) and octupole ($E3$) transition 
probabilities are analyzed. 
The $E\lambda$ ($\lambda=2,3$) operator 
is composed of the boson and
fermion contributions: 
\begin{eqnarray}
 \hat T^{(E\lambda)} = \hat T^{(E\lambda)}_{\rm B} + \hat T^{(E\lambda)}_{\rm F} \; .
\end{eqnarray}
For the $E2$ operator, the bosonic part reads 
$\hat T^{(E2)}_{\rm B}=e_{\rm B}^{(2)}\hat Q$, with the quadrupole operator  
$\hat Q$ defined in Eq.~(\ref{eq:quad}), and the fermion $E2$
operator 
\begin{eqnarray}
 \hat T^{(E2)}_{\rm F}=
-e_{\rm F}^{(2)}\sum_{j_aj_b}
\frac{1}{\sqrt{5}}
\gamma_{j_a,j_b}^{(2)}
(a^{\+}_{j_a}\times\tilde a_{j_b})^{(2)}. 
\end{eqnarray}
$e^{(2)}_{\rm B}$ and $e^{(2)}_{\rm F}$ denote 
bosonic and fermion E2 effective
charges, respectively. 
Here $e^{(2)}_{\rm B}$ is determined 
for each nucleus so as to reproduce the 
experimental $B(E2; 2^+_1\rightarrow 0^+_1)$ 
value for the even-even Xe core \cite{data}, 
and the employed boson charges are 
$e^{(2)}_{\rm B}=0.12$ $e$b (for $^{140,141}$Xe), 
0.158 $e$b (for $^{142,143}$Xe), 
and 0.106 $e$b (for $^{144,145}$Xe). 
The fermionic effective charge 
$e^{(2)}_{\rm F}=0.5$ $e$b is used for all the 
considered odd-mass Xe isotopes. 
Similarly, for the $E3$ transition operator, 
the bosonic part reads 
$\hat T^{(E3)}=e^{(3)}_{\rm B} \hat O$, and the fermion part
can be written, in analogy to the quadrupole 
one, as
\begin{eqnarray}
 \hat T^{(E3)}_{\rm F} = -e_{\rm F}^{(3)}\sum_{j_aj_b'}
\frac{1}{\sqrt{7}}
\gamma_{j_a,j_b'}^{(3)}
(a^{\+}_{j_a}\times\tilde a_{j_b'})^{(3)}. 
\end{eqnarray}
Following Ref.~\cite{nomura2021oct-ba}, 
the $E3$ boson effective charge 
$e^{(3)}_{\rm B}$ is assumed to be a function of 
those $\beta_2$ and $\beta_3$ deformations 
corresponding to the global minimum, 
$e^{(3)}_{\rm B} \propto (1+\bar\beta_2 \bar\beta_3)$ $e$b$^{3/2}$, 
while the fixed $E3$ fermion charge of 
$e^{(3)}_{\rm F}=0.5\,eb^{3/2}$ 
is employed. 
The adopted $e^{(3)}_{\rm B}$ values 
are $0.14$ $e$b$^{3/2}$ (for $^{140,141}$Xe), 
$0.199$ $e$b$^{3/2}$ (for $^{142,143}$Xe), and 
$0.14$ $e$b$^{3/2}$ (for $^{144,145}$Xe), 
which are determined so that the 
$B(E3; 3^-_1\rightarrow 0^+_1)$ rates 
for the even-even nuclei 
are close to and exhibit similar 
behaviors as functions of $N$ to 
those obtained in the previous $sdf$-IBM 
study for the neutron-rich Xe, Ba, Ce, and Nd 
nuclei based on the Gogny force \cite{nomura2021oct-ba}.

%
%
\begin{figure}[ht]
\begin{center}
\includegraphics[width=.8\linewidth]{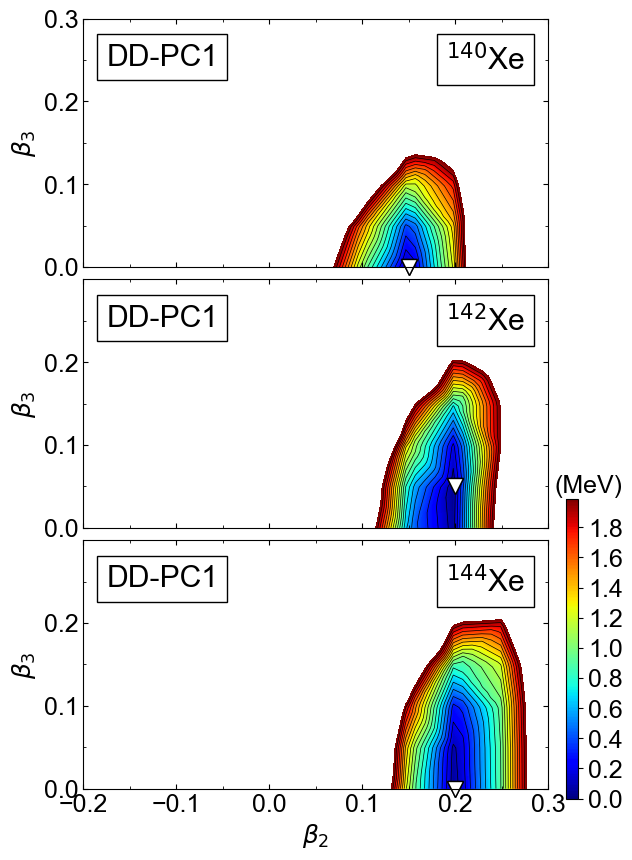}
\caption{Axially symmetric 
quadrupole ($\beta_2$) - octupole ($\beta_3$)
constrained PESs for the even-even 
$^{140,142,144}$Xe isotopes, computed by the 
RHB method using the DD-PC1 functional and a 
separable pairing force of finite range. 
Total SCMF energies are plotted up to 
2 MeV with respect to the global minimum, 
indicated by the open triangle. 
Energy difference between neighboring contours 
is 100 keV.}
\label{fig:pesdft}
\end{center}
\end{figure}

%
%
\begin{figure}[ht]
\begin{center}
\includegraphics[width=.8\linewidth]{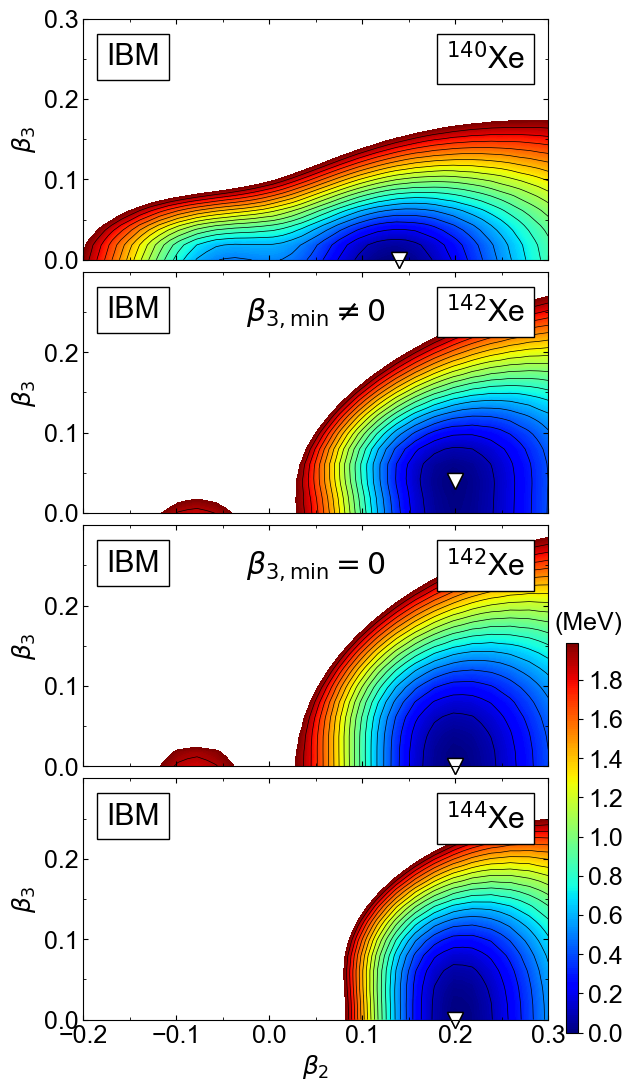}
\caption{
Similar to the caption to 
Fig.~\ref{fig:pesdft}, but for the mapped $sdf$-IBM PESs.
Two different PESs for the $^{142}$Xe are obtained 
by taking $\beta_3=0.05$ ($\neq 0$) as the global minimum 
consistent with the corresponding RHB-SCMF PES 
in Fig.~\ref{fig:pesdft}, and by assuming that the 
global minimum occurs at $\beta_3=0$.}
\label{fig:pesibm}
\end{center}
\end{figure}

%
%
\begin{figure}[htb!]
\begin{center}
\includegraphics[width=\linewidth]{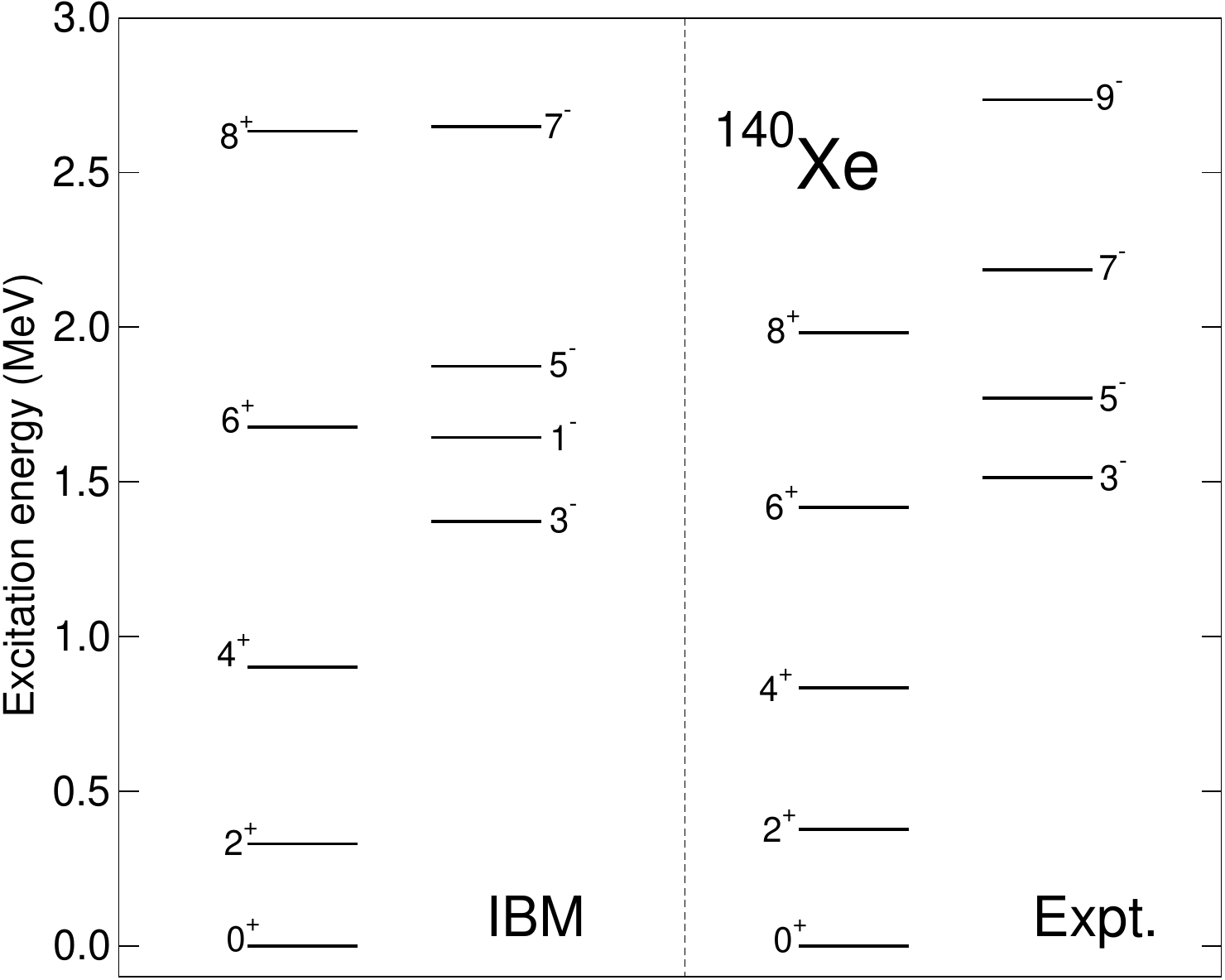}\\
\includegraphics[width=\linewidth]{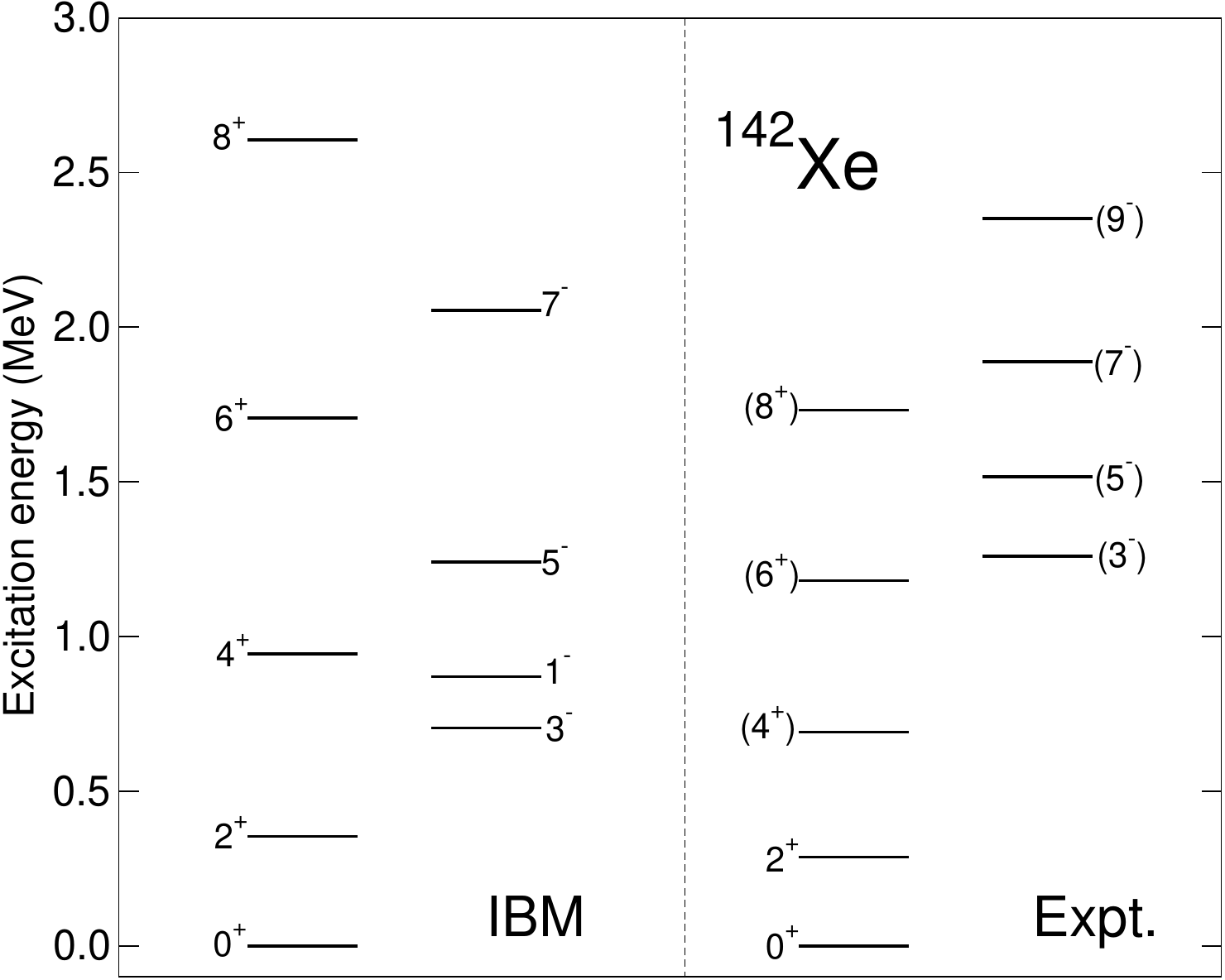}\\
\includegraphics[width=\linewidth]{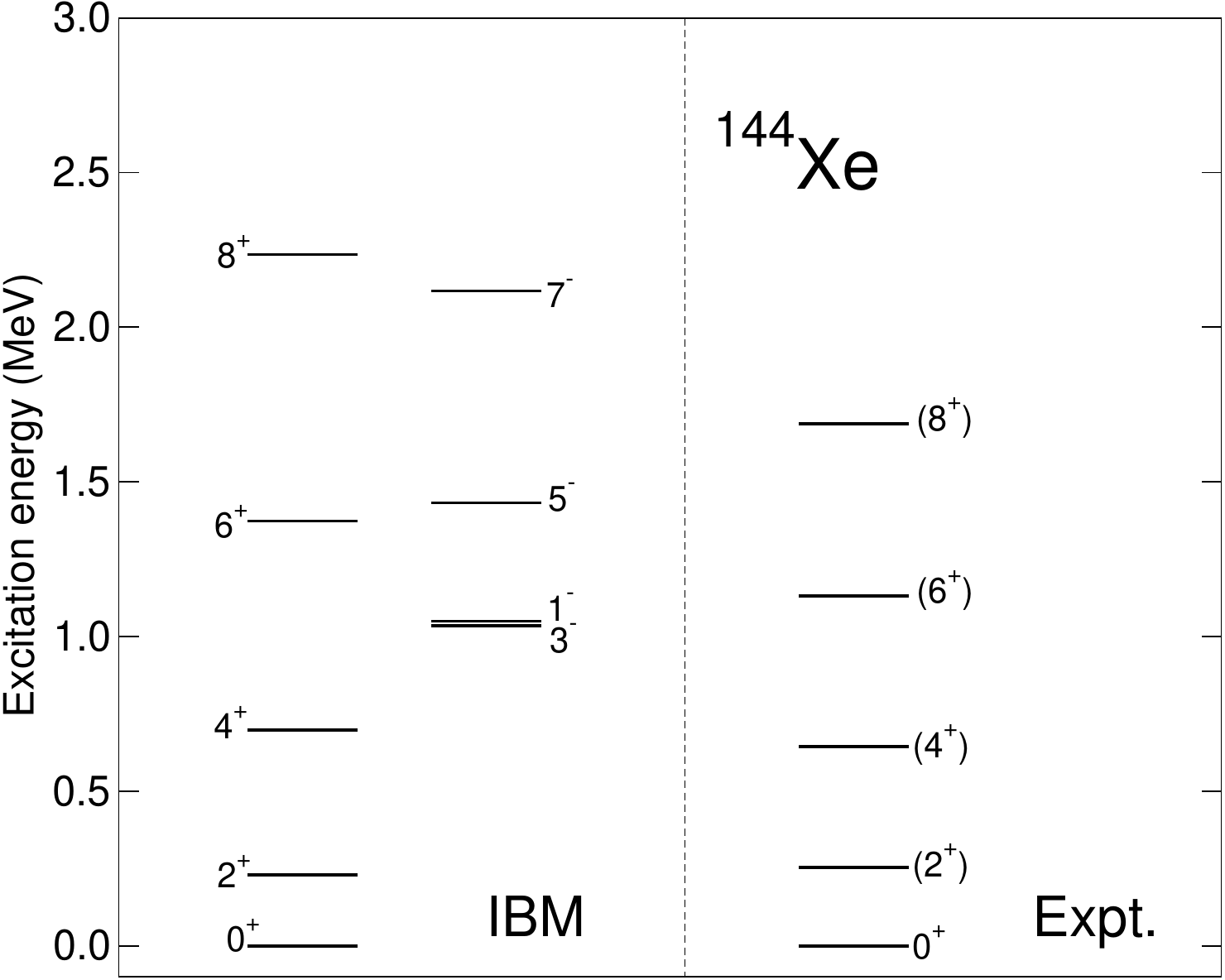}
\caption{Calculated low-energy positive- and negative-parity 
spectra for the even-even $^{140,142,144}$Xe nuclei. 
Experimental data are taken from the ENSDF database \cite{data}.} 
\label{fig:evenxe}
\end{center}
\end{figure}

%
%
\begin{figure}[htb!]
\begin{center}
\includegraphics[width=\linewidth]{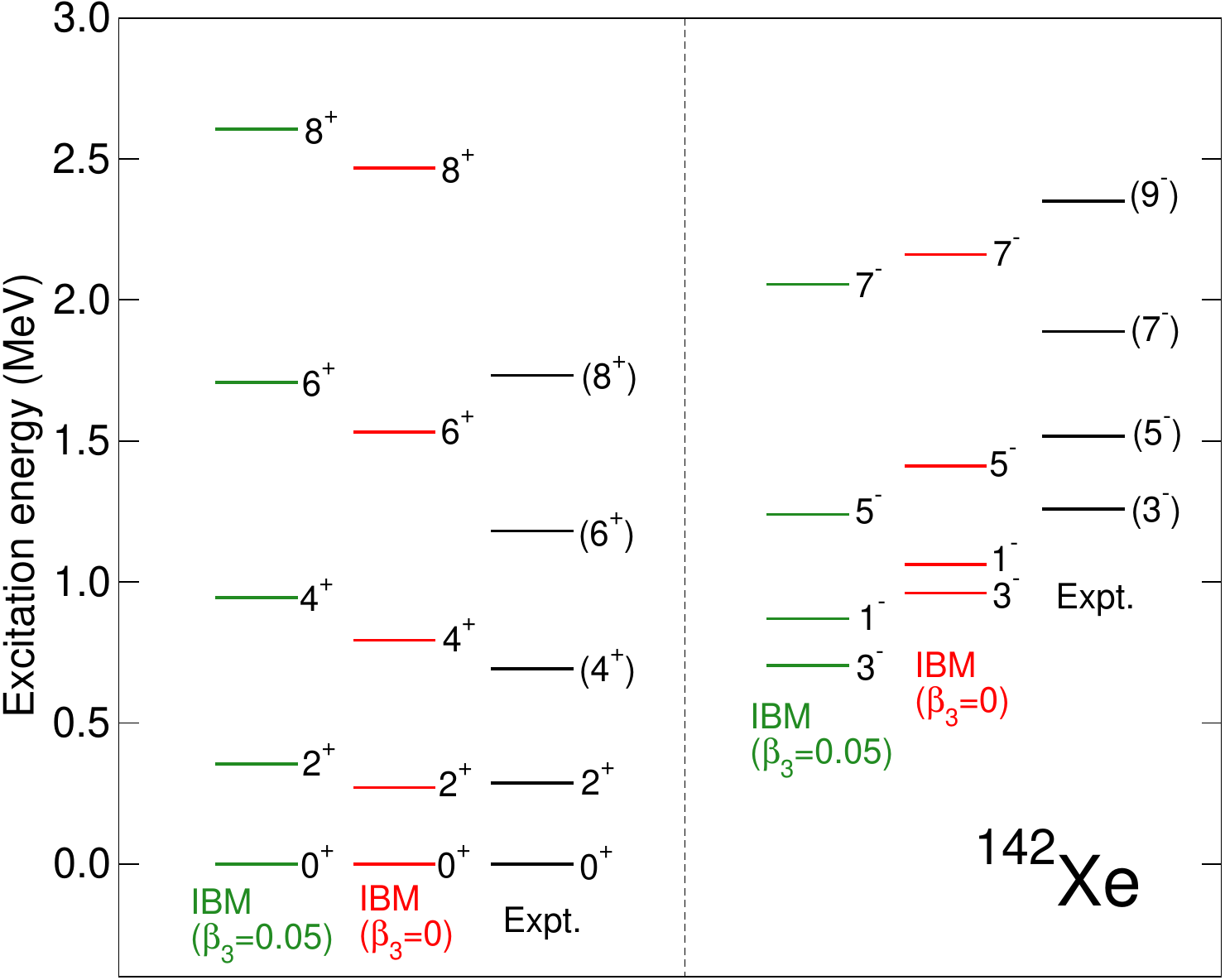}
\caption{Same as Fig.~\ref{fig:evenxe}, but for the 
calculated energy spectra for the 
$^{142}$Xe nucleus obtained by taking 
the $\beta_3=0.05(\neq 0)$ global minimum and by assuming 
that the minimum occurs at $\beta_3=0$.} 
\label{fig:142soft}
\end{center}
\end{figure}

\section{Results for even-even Xe\label{sec:even}}

Figure~\ref{fig:pesdft} shows the RHB-SCMF PESs 
in terms of the axially symmetric 
quadrupole $\beta_2$ and octupole $\beta_3$ 
deformations for the 
even-even nuclei $^{140,142,144}$Xe, plotted up to the 
2 MeV from the global minimum. 
All these nuclei exhibit prolate deformation along the 
$\beta_2$ deformation with the equilibrium minimum 
found at $\beta_2 \approx 0.15$, 0.20, and 0.20 for 
$^{140}$Xe, $^{142}$Xe, $^{144}$Xe, respectively. 
Notable features are that for $^{142}$Xe ($N=88$) 
the global minimum 
is found at a non-zero octupole deformation, 
$(\beta_2,\beta_3)=(0.2,0.05)$, 
and that the potential is soft 
along $\beta_3$ deformation, as the 
energy at the global minimum is 
only 32 keV lower than the lowest energy along the 
$\beta_3=0$ axis. 
These characteristics of the PESs indicate 
the relevance of the octupole correlations 
in the $^{142}$Xe nucleus, which also confirms 
that the octupole deformation is most pronounced 
at $N=88$. 
In the previous study of Ref.~\cite{nomura2018oct}, 
the RHB-SCMF calculation 
for the $^{144}$Ba nucleus also suggested a $\beta_3$-soft 
potential but with a much steeper 
$\beta_3\neq 0$ global minimum at 
$(\beta_2,\beta_3) \approx (0.2,0.1)$. 
For $^{144}$Xe, 
the non-zero octupole minimum 
is no longer seen, but the potential is still soft in the 
$\beta_3$ deformation, to a similar extent to that 
for the $^{142}$Xe.

The mapped $sdf$-IBM PESs are plotted 
in Fig.~\ref{fig:pesibm}. 
They look rather soft particularly with respect to 
the $\beta_2$ deformation as compared to the RHB-SCMF 
counterparts. This is a consequence of the fact that the 
mapping is made primarily to reproduce 
the topology of the RHB-SCMF PESs in the vicinity 
of the global minimum, specifically, 
the softness in the $\beta_3$ deformation 
and the location of the global minimum. 
As for $^{142}$Xe, in order to see impacts of 
the shallow $\beta_3\neq 0$ minimum on energy spectra, 
another set of the $sdf$-IBM 
and $sdf$-IBFM calculations is carried out 
with an assumption that the minimum occurs 
at $\beta_3=0$, instead of $\beta_3=0.05$. 
The IBM PES obtained with such an assumption 
is also plotted in Fig.~\ref{fig:pesibm}. 
The derived $sdf$-IBM parameters with the 
$\beta_3=0$ global minimum slightly 
differ from those 
with the calculation with the non-zero 
octupole deformation, such that 
$\epsilon_f=0.851$ MeV and $\chi'=-2.0$.

Figure~\ref{fig:evenxe} displays 
the calculated low-energy positive- and negative-parity spectra 
for the even-even $^{140,142,144}$Xe isotopes, 
obtained from the diagonalization of the mapped $sdf$-IBM 
Hamiltonian (\ref{eq:bh}) with parameters determined 
from the RHB-SCMF calculations. 
The results are compared 
with the experimental data available 
at ENSDF database of NNDC \cite{data}. 
The calculation reasonably reproduces the $2^+_1$ 
energies, but the energy levels in the 
ground-state bands are rather overestimated 
for higher-spin states with $I^\pi > 4^+$. 
This is partly due to the 
limited number of bosons considered in the $sdf$-IBM, 
i.e., $n=4$, 5, and 6 for 
$^{140,142,144}$Xe, respectively, which may be 
insufficient to account for the energies 
for the states higher than $I^\pi=4^+$. 
For all the even-even Xe nuclei, the $sdf$-IBM 
predicts the lowest negative-parity state to be 
$I^\pi=3^-$, above which the $1^-$ level appears. 
In the previous study of Ref.~\cite{nomura2018oct} 
a similar level structure of negative-parity odd-spin 
states to those of $^{140,142,144}$Xe, shown in 
Fig.~\ref{fig:evenxe}, were obtained for $^{142}$Ba, 
where octupole deformation is minor. 
For $^{144,146}$Ba, on the other hand, 
the $sdf$-IBM predicted a rather rotational-like negative-parity 
yrast band, with the bandhead being the $1^-$ state \cite{nomura2018oct}. 
For $^{140}$Xe the present $sdf$-IBM calculation 
reproduces the $3^-$ energy reasonably well, 
while the higher-spin negative-parity levels are 
overestimated. 
For $^{142}$Xe, the mapped $sdf$-IBM gives much lower 
$3^-$ and $5^-$ energy levels than the experimental 
values, which reflects the pronounced $\beta_3$ 
softness with $\beta_3\neq 0$ global minimum in 
its PES (see Fig.~\ref{fig:pesdft}).

Figure~\ref{fig:142soft} compares between the 
$sdf$-IBM energy spectra obtained 
with and without taking into account 
the non-zero octupole minimum in the mapping procedure. 
A notable difference between the two $sdf$-IBM results 
is that the negative-parity levels obtained 
with the $\beta_3=0$ minimum are higher 
in energy than those obtained with $\beta_3\neq 0$, 
and this difference indicates that the octupole effects 
taken into account in the $sdf$-IBM are less significant 
in the former calculation than in the latter. 

In Table~\ref{tab:frac-even} expectation values of the 
$f$-boson number operator $\braket{\hat n_f}$ 
calculated for the positive- and negative-parity 
yrast states are presented. 
For $^{140}$Xe, 
$f$ bosons appear to play a minor role in the 
positive-parity states, since the 
expectation value $\braket{\hat n_f} < 0.1$, whereas 
the negative-parity states are described 
by one-$f$-boson configurations. 
In $^{142}$Xe,  the $f$-boson contributions are 
suggested to be significant even in the 
ground state, $\braket{\hat n_f} > 1$. 
This finding is similar to that 
of the previous mapped $sdf$-IBM calculation 
for the neutron-rich even-even Xe nuclei based 
on the Gogny EDF \cite{nomura2021oct-ba}. 
Table~\ref{tab:frac-even} shows that 
large amounts of the $f$-boson components 
are present also in the calculated 
states for $^{144}$Xe.

\begin{table}[hb!]
\caption{\label{tab:frac-even}
Expectation values of the $f$-boson 
number operator $\braket{\hat n_f}$ for 
the yrast states for $^{140,142,144}$Xe. 
}
\begin{center}
 \begin{ruledtabular}
\begin{tabular}{cccc}
\textrm{$I^\pi$} &
\textrm{$^{140}$Xe} &
\textrm{$^{142}$Xe} &
\textrm{$^{144}$Xe} \\
\hline
$ {0}^{+}_{1}$ & $0.058$ & 1.121 & 0.575 \\ 
$ {2}^{+}_{1}$ & $0.031$ & 0.818 & 0.467 \\ 
$ {4}^{+}_{1}$ & $0.014$ & 0.585 & 0.353 \\ 
$ {6}^{+}_{1}$ & $0.005$ & 0.772 & 0.253 \\ 
$ {8}^{+}_{1}$ & $0.001$ & 1.508 & 0.172 \\ 
$ {1}^{-}_{1}$ & $1.019$ & 1.434 & 1.366 \\ 
$ {3}^{-}_{1}$ & $1.046$ & 1.714 & 1.460 \\ 
$ {5}^{-}_{1}$ & $1.017$ & 1.432 & 1.344 \\ 
$ {7}^{-}_{1}$ & $1.004$ & 1.241 & 1.229 \\ 
\end{tabular}
 \end{ruledtabular}
\end{center}
\end{table}

\begin{table}[hb!]
\caption{\label{tab:tr-even}
$B(E2)$ and $B(E3)$ transition rates in 
Weisskopf units (W.u.) for the 
low-lying states of the even-even $^{140,142,144}$Xe nuclei 
calculated with the mapped $sdf$-IBM, and 
the experimental data available at the NNDC \cite{data}.}
\begin{center}
 \begin{ruledtabular}
\begin{tabular}{cccc}
\textrm{$A$} &
\textrm{$B(E\lambda;I^{\pi}_i \to I^{\pi}_f)$} &
\textrm{$sdf$-IBM (W.u.)}&
\textrm{Expt. (W.u.)} \\
\hline
$^{140}$Xe
& $ B(E2; {2}^{+}_{1} \to {0}^{+}_{1})$ & $24.1$ & $24.0\pm0.7$ \\ 
& $ B(E2; {4}^{+}_{1} \to {2}^{+}_{1})$ & $32.5$ & $45^{+9}_{-6}$ \\ 
& $ B(E2; {6}^{+}_{1} \to {4}^{+}_{1})$ & $28.6$ & $>22.6$ \\ 
& $ B(E2; {5}^{-}_{1} \to {3}^{-}_{1})$ & $16.5$ & $64^{+38}_{-22}$ \\ 
& $ B(E3; {3}^{-}_{1} \to {0}^{+}_{1})$ & $34.7$ & ${}$ \\ 
& $ B(E3; {5}^{-}_{1} \to {2}^{+}_{1})$ & $41.0$ & ${}$ \\ 
& $ B(E3; {7}^{-}_{1} \to {4}^{+}_{1})$ & $34.0$ & ${}$ \\ 
$^{142}$Xe
& $ B(E2; {2}^{+}_{1} \to {0}^{+}_{1})$ & $31.4$ & $31.4\pm4.5$ \\ 
& $ B(E2; {4}^{+}_{1} \to {2}^{+}_{1})$ & $52.5$ & ${}$ \\ 
& $ B(E2; {6}^{+}_{1} \to {4}^{+}_{1})$ & $48.0$ & ${}$ \\ 
& $ B(E3; {3}^{-}_{1} \to {0}^{+}_{1})$ & $21.5$ & ${}$ \\ 
& $ B(E3; {5}^{-}_{1} \to {2}^{+}_{1})$ & $40.7$ & ${}$ \\ 
& $ B(E3; {7}^{-}_{1} \to {4}^{+}_{1})$ & $53.3$ & ${}$ \\ 
$^{144}$Xe
& $ B(E2; {2}^{+}_{1} \to {0}^{+}_{1})$ & $32.6$ & $32.6\pm7.6$ \\ 
& $ B(E2; {4}^{+}_{1} \to {2}^{+}_{1})$ & $46.7$ & ${}$ \\ 
& $ B(E2; {6}^{+}_{1} \to {4}^{+}_{1})$ & $48.6$ & ${}$ \\ 
& $ B(E3; {3}^{-}_{1} \to {0}^{+}_{1})$ & $20.9$ & ${}$ \\ 
& $ B(E3; {5}^{-}_{1} \to {2}^{+}_{1})$ & $33.7$ & ${}$ \\ 
& $ B(E3; {7}^{-}_{1} \to {4}^{+}_{1})$ & $39.6$ & ${}$ \\
\end{tabular}
 \end{ruledtabular}
\end{center}
\end{table}

Table~\ref{tab:tr-even} lists predicted 
$B(E2)$ and $B(E3)$ transition rates for the 
low-spin positive- and negative-parity states 
of the even-even $^{140,142,144}$Xe nuclei, 
in comparison with the experimental data \cite{data}. 
There is no experimental data available for 
$^{142,144}$Xe, other than the 
$B(E2;2^+_1 \to 0^+_1)$ rates \cite{data}. 
The present mapped $sdf$-IBM calculation 
predicts in-band $E2$ transitions within the 
ground-state band. The predicted
$B(E2;5^-_1 \to 3^-_1)$ transition probability 
for $^{140}$Xe underestimates the experimental value. 
The calculation suggests 
sizable $E3$ transitions of 
$B(E3;3^-_1 \to 0^+_1)\approx 20-30$ Weisskopf units (W.u.) 
for the studied even-even Xe isotopes.

%
%
\begin{figure}[htb!]
\begin{center}
\includegraphics[width=\linewidth]{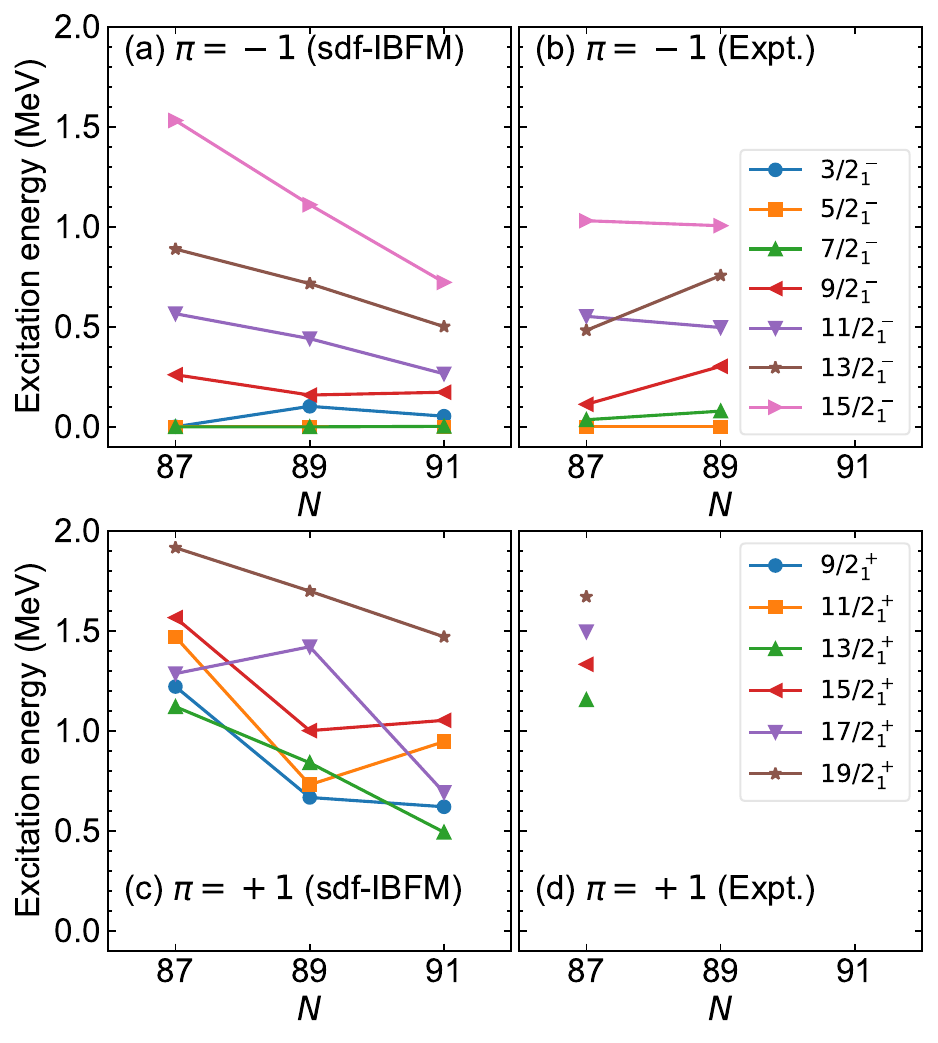}
\caption{Evolution of low-lying negative-parity 
[$\pi=-1$, (a) and (b)] 
and positive-parity 
[$\pi=+1$, (c) and (d)] 
spectra for the 
odd-mass nuclei $^{141,143,145}$Xe as functions 
of $N$. The calculated 
results from the $sdf$-IBFM are shown in (a) and (c), 
and the experimental data \cite{data} 
are depicted in (b) and (d).} 
\label{fig:oddxe}
\end{center}
\end{figure}

%
%
\begin{figure*}[htb!]
\begin{center}
\includegraphics[width=.8\linewidth]{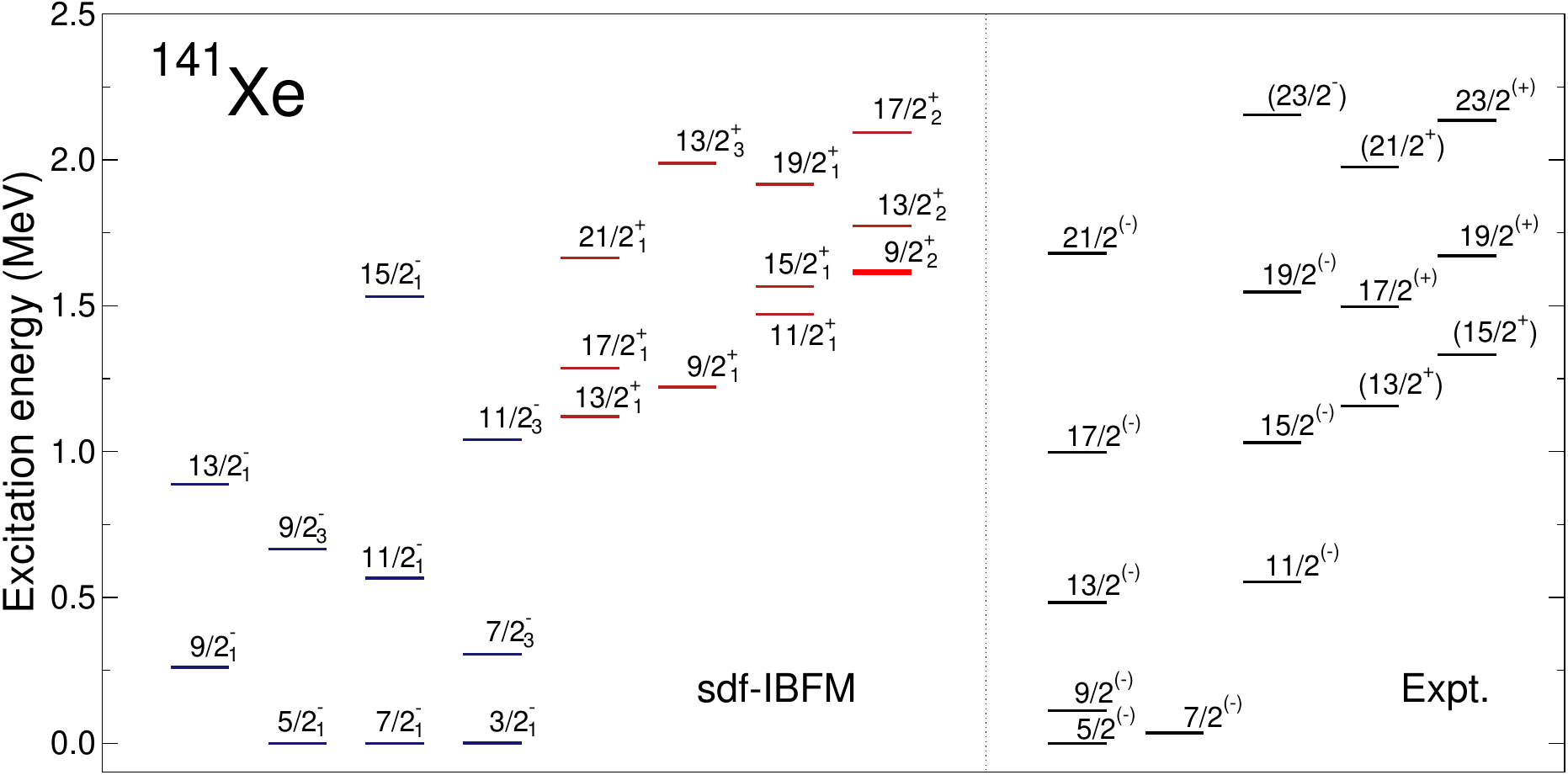}
\caption{Calculated and experimental \cite{data} 
low-energy spectra for $^{141}$Xe. 
Those calculated states that contain more than one 
$f$-bosons in the wave functions, $\braket{\hat n_f}>1$, 
are highlighted in thick lines.} 
\label{fig:xe141}
\end{center}
\end{figure*}

%
%
\begin{figure*}[htb!]
\begin{center}
\includegraphics[width=.8\linewidth]{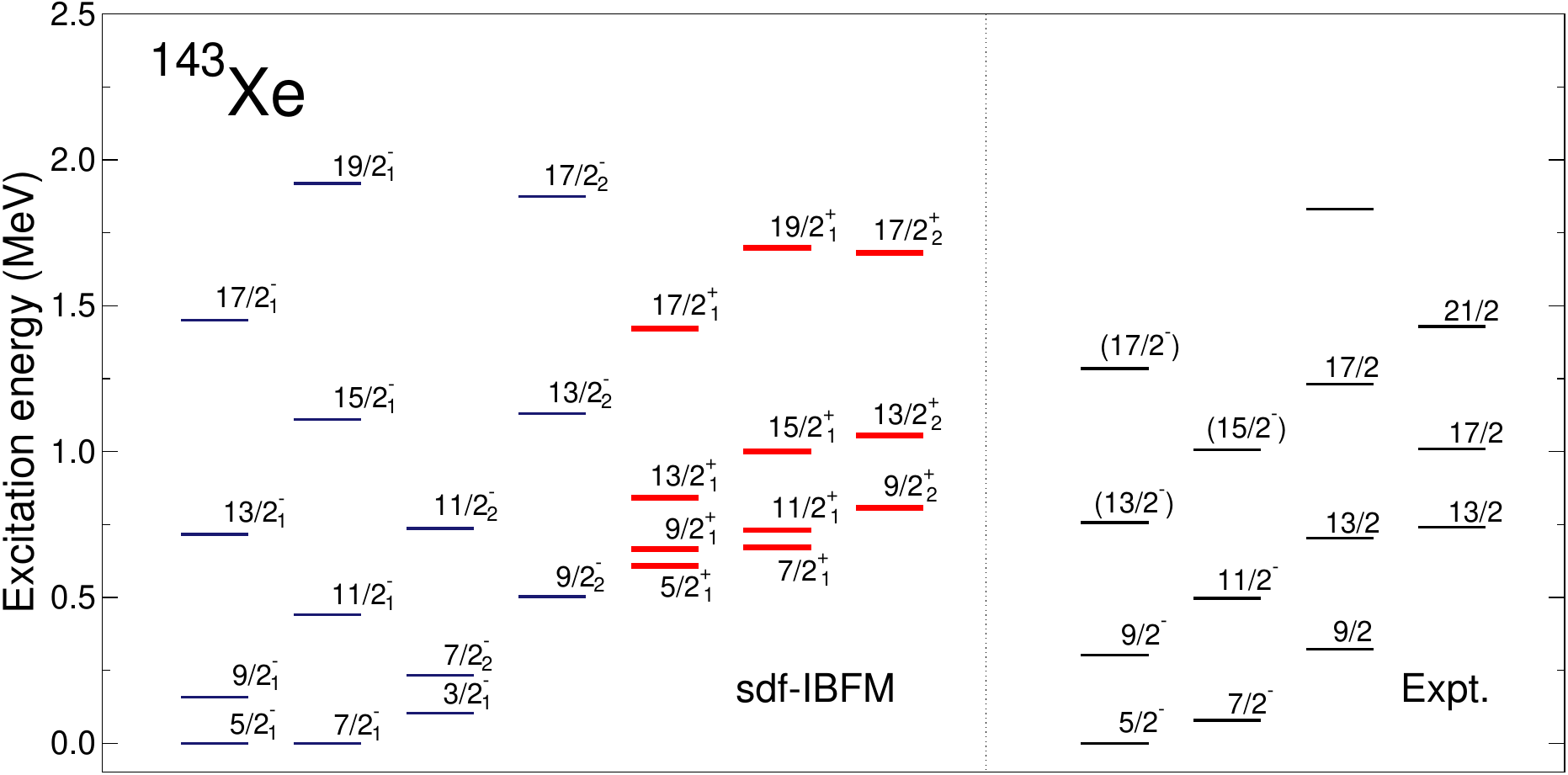}
\caption{Same as Fig.~\ref{fig:xe141}, but for $^{143}$Xe. 
Experimental data 
are taken from Ref.~\cite{rzacaurban2011}.} 
\label{fig:xe143}
\end{center}
\end{figure*}

%
%
\begin{figure}[htb!]
\begin{center}
\includegraphics[width=\linewidth]{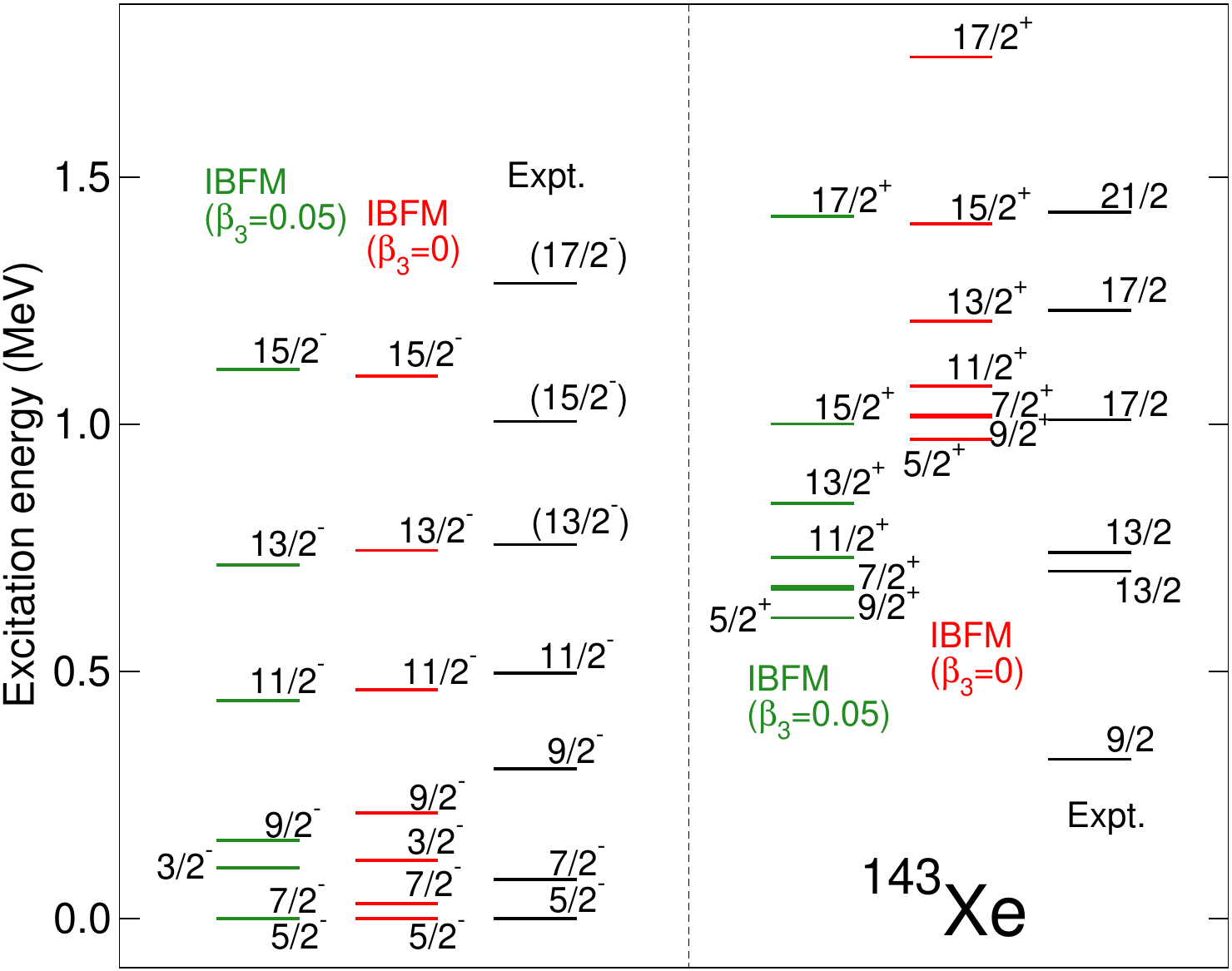}
\caption{Calculated negative- (left) and positive-parity (right) 
energy spectra for the $^{143}$Xe 
nucleus obtained with the non-zero octupole 
($\beta_3=0.05\neq 0$) global minimum and with the 
assumption of the zero-octupole ($\beta_3=0$) minimum 
in the mapping. Experimental data are taken from 
Ref.~\cite{rzacaurban2011}. Note that the parity 
of the experimental levels 
shown on the right-hand side has not been determined.} 
\label{fig:143soft}
\end{center}
\end{figure}

%
%
\begin{figure}[htb!]
\begin{center}
\includegraphics[width=\linewidth]{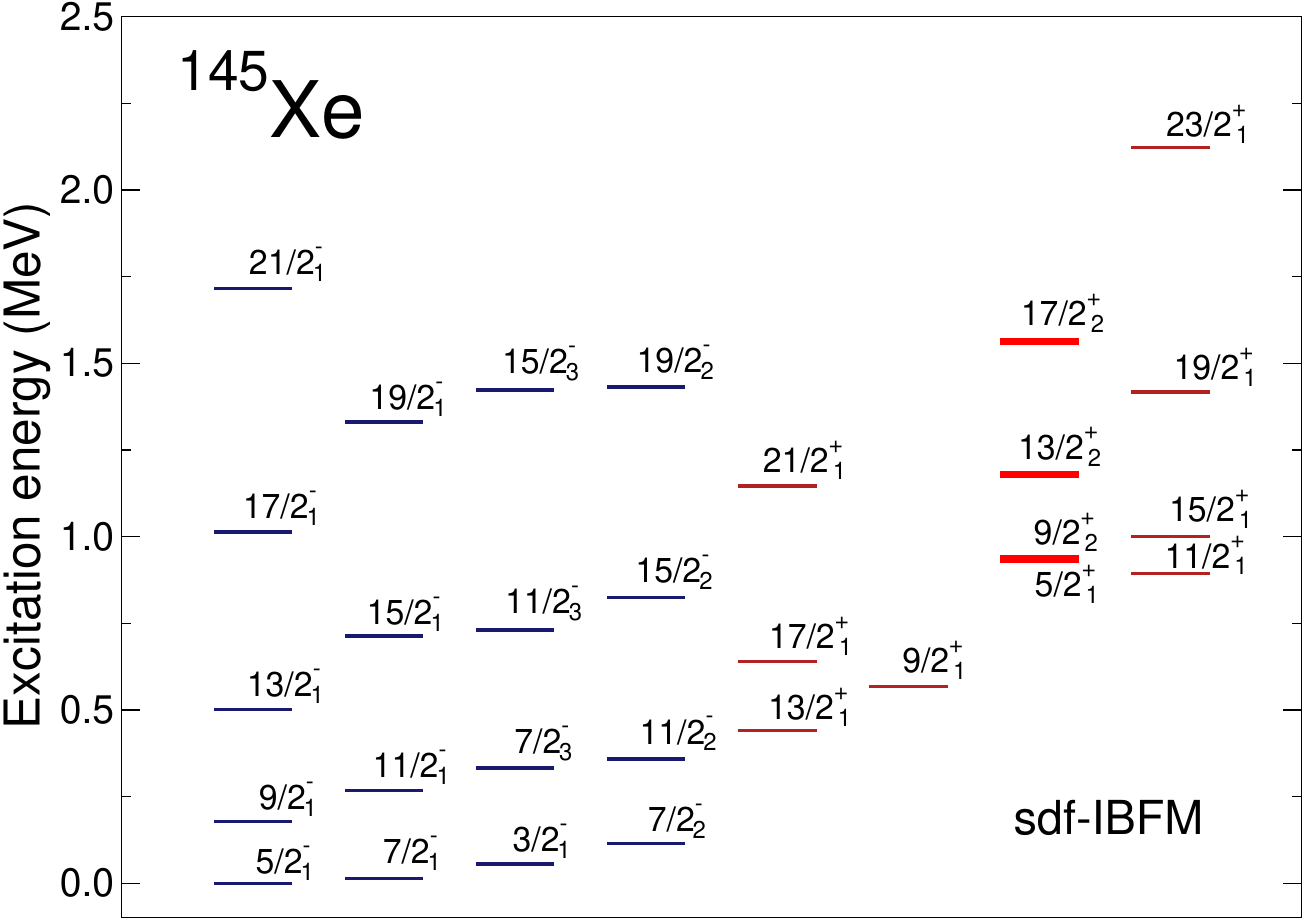}
\caption{Calculated low-energy spectra for $^{145}$Xe.} 
\label{fig:xe145}
\end{center}
\end{figure}

\section{Results for odd-mass Xe\label{sec:odd}}

\subsection{Evolution of energy levels}

In Fig.~\ref{fig:oddxe} calculated low-lying positive- 
and negative-parity energy spectra for the odd-mass 
$^{141,143,145}$Xe nuclei are depicted 
as functions of $N$, 
obtained from the $sdf$-IBFM with a microscopic input 
provided by the RHB-SCMF calculation using 
the DD-PC1 EDF. 
Experimental spectra 
adopted from the NNDC \cite{data} are also plotted. 
The present calculation suggests that 
the negative-parity states are composed mainly 
of the configurations of a single neutron in 
the normal-parity ($pfh$) orbitals coupled to 
the $sd$-boson space. 
The decreasing pattern of the calculated 
negative-parity levels reflects evolution of 
the quadrupole collectivity with 
increasing valence nucleon number. 
The behaviors of the calculated 
${9/2}^-_1$ and ${13/2}^-_1$ levels from 
$N=87$ to 89 are, however, at variance with 
the experimental ones which 
rather increase with $N$. 
A reasonable agreement with the observed 
negative-parity levels is obtained for $^{143}$Xe.

In Fig.~\ref{fig:oddxe}(c), one can observe that 
the structure of the calculated positive-parity levels 
for $^{143}$Xe is rather different from those of 
the adjacent nuclei $^{141,145}$Xe, namely, 
at $N=89$ the ${9/2}^+_1$ and ${11/2}^+_1$ levels  
become particularly low in energy such that 
they are below the ${13/2}^+_1$ level. 
The change in the nuclear structure at $N=89$ 
correlates with the fact that for $^{142}$Xe 
a potential is rather soft and exhibits 
a non-zero $\beta_3$ minimum. 
Such a feature of the PES is supposed to be reflected 
in some of those derived $sdf$-IBM parameters for $^{142}$Xe
that are relevant to the octupole degree of freedom, 
and hence in the fitted 
boson-fermion interaction strengths for $^{143}$Xe. 
For example, as shown in Table~\ref{tab:paraB} 
the single-$f$ boson energy $\epsilon_f$ is rather low 
at $^{142}$Xe as compared to those derived for $^{140,144}$Xe, 
and the largest value of the 
parameter $\chi'$ is chosen for $^{142}$Xe 
among the three even-even Xe nuclei. 
Also in Table~\ref{tab:paraBF} the parameter 
$\Lambda_0^{ff}=0.60$ MeV,  
employed for the positive-parity states of $^{143}$Xe, 
is much larger than the corresponding values 
for $^{141,145}$Xe. 
As will be shown, many of the low-energy positive-parity 
states for $^{143}$Xe are accounted for by large 
amounts of octupole-boson contributions, whereas 
in $^{141}$Xe the dominant configurations in these 
positive-parity states are based on 
the neutron $\nu 1i_{13/2}$ single-particle orbital coupled 
to the $sd$-boson space.

\subsection{Detailed energy spectrum}

Figures~\ref{fig:xe141}--\ref{fig:xe145} 
depict detailed energy spectra resulting 
from the $sdf$-IBFM calculations. 
The calculated energy levels shown in 
Figs.~\ref{fig:xe141}, \ref{fig:xe143}, and \ref{fig:xe145} 
are grouped into bands so as to follow dominant $\Delta I=2$ $E2$ 
transitions between states and the increasing 
order of the angular momenta. 
The experimental data are taken from the NNDC \cite{data}.

\subsubsection{$^{141}$Xe}

%
%
\begin{table}[htb!]
\caption{\label{tab:frac-xe141}
Expectation values of the $f$-boson number 
operator $\hat n_f$, and the single-particle 
operators $\hat n_j$ corresponding to each orbital 
in those states for the $^{141}$Xe nucleus shown 
in Fig.~\ref{fig:xe141}.}
\begin{center}
\begin{ruledtabular}
\begin{tabular}{cccccccc}
\multirow{2}{*}{$I^{\pi}$} &
\multirow{2}{*}{$\braket{\hat n_f}$} &
\multicolumn{6}{c}{$\braket{n_j}$} \\
\cline{3-8}
& &
\textrm{$3p_{1/2}$} &
\textrm{$3p_{3/2}$} &
\textrm{$2f_{5/2}$} &
\textrm{$2f_{7/2}$} &
\textrm{$1h_{9/2}$} &
\textrm{$1i_{13/2}$} \\
\hline
$ {3/2}^{-}_{1}$ & $0.012$ & $0.006$ & $0.039$ & $0.018$ & $0.657$ & $0.279$ & $0.002$ \\ 
$ {5/2}^{-}_{1}$ & $0.009$ & $0.006$ & $0.001$ & $0.046$ & $0.030$ & $0.916$ & $0.000$ \\ 
$ {7/2}^{-}_{1}$ & $0.008$ & $0.002$ & $0.007$ & $0.020$ & $0.248$ & $0.722$ & $0.000$ \\ 
$ {7/2}^{-}_{3}$ & $0.015$ & $0.004$ & $0.014$ & $0.024$ & $0.377$ & $0.581$ & $0.001$ \\ 
$ {9/2}^{-}_{1}$ & $0.003$ & $0.006$ & $0.014$ & $0.021$ & $0.344$ & $0.615$ & $0.000$ \\ 
$ {9/2}^{-}_{2}$ & $0.002$ & $0.011$ & $0.014$ & $0.022$ & $0.385$ & $0.567$ & $0.000$ \\ 
$ {11/2}^{-}_{1}$ & $0.001$ & $0.008$ & $0.015$ & $0.022$ & $0.413$ & $0.543$ & $0.000$ \\ 
$ {11/2}^{-}_{3}$ & $0.005$ & $0.006$ & $0.008$ & $0.018$ & $0.381$ & $0.588$ & $0.000$ \\ 
$ {13/2}^{-}_{1}$ & $0.000$ & $0.015$ & $0.026$ & $0.019$ & $0.412$ & $0.527$ & $0.000$ \\ 
$ {13/2}^{-}_{3}$ & $0.001$ & $0.000$ & $0.035$ & $0.080$ & $0.520$ & $0.364$ & $0.000$ \\ 
$ {15/2}^{-}_{1}$ & $0.001$ & $0.000$ & $0.050$ & $0.063$ & $0.310$ & $0.576$ & $0.000$ \\ 
$ {9/2}^{+}_{1}$ & $0.033$ & $0.000$ & $0.000$ & $0.000$ & $0.013$ & $0.001$ & $0.985$ \\ 
$ {9/2}^{+}_{2}$ & $1.003$ & $0.003$ & $0.005$ & $0.022$ & $0.134$ & $0.833$ & $0.004$ \\ 
$ {11/2}^{+}_{1}$ & $0.040$ & $0.000$ & $0.000$ & $0.000$ & $0.018$ & $0.003$ & $0.978$ \\ 
$ {13/2}^{+}_{1}$ & $0.033$ & $0.000$ & $0.000$ & $0.000$ & $0.004$ & $0.000$ & $0.996$ \\ 
$ {13/2}^{+}_{2}$ & $0.038$ & $0.000$ & $0.000$ & $0.000$ & $0.014$ & $0.002$ & $0.984$ \\ 
$ {15/2}^{+}_{1}$ & $0.023$ & $0.000$ & $0.000$ & $0.000$ & $0.004$ & $0.000$ & $0.996$ \\ 
$ {17/2}^{+}_{1}$ & $0.021$ & $0.000$ & $0.000$ & $0.000$ & $0.002$ & $0.000$ & $0.998$ \\ 
$ {17/2}^{+}_{2}$ & $0.015$ & $0.000$ & $0.000$ & $0.000$ & $0.004$ & $0.000$ & $0.996$ \\ 
$ {19/2}^{+}_{1}$ & $0.011$ & $0.000$ & $0.000$ & $0.000$ & $0.001$ & $0.000$ & $0.998$ \\ 
$ {21/2}^{+}_{1}$ & $0.010$ & $0.000$ & $0.000$ & $0.000$ & $0.001$ & $0.000$ & $0.999$ \\ 
\end{tabular}
\end{ruledtabular}
\end{center}
\end{table}

The even-even core nucleus $^{140}$Xe is here suggested 
to be mostly quadrupole deformed, and hence the 
octupole effects are also expected to be weak in the 
odd-mass neighbor $^{141}$Xe. 
In the experimental studies 
of Refs.~\cite{urban2000-141Xe,huang2017-141Xe} 
two pairs of $\Delta I=2$ bands, 
one consisting of the ${5/2}^-_1$ and ${15/2}^+_1$ bands, 
and the other of the ${11/2}^-_1$ and ${13/2}^+_1$ bands 
were interpreted as simplex symmetry partners 
with $s=\mp i$, respectively. 
As shown in Fig.~\ref{fig:xe141} negative-parity bands 
built on the ${3/2}^+_1$, ${5/2}^+_1$, ${7/2}^+_1$, 
and ${9/2}^+_1$ states are suggested by the $sdf$-IBFM. 
These bands are connected by only weak $E2$ transitions, 
which are typically less than 10 W.u., and are 
uniformly stretched in energy 
in comparison to the observed ones. 

The present calculation, however, gives
the $\Delta I =2$ positive-parity band built on
the ${13/2}^+_1$ level, with the 
band-head level in agreement with experiment. 
The predicted ${13/2}^+_1$ band also exhibits 
large $E2$ transitions: 
$B(E2;{17/2}^+_1\to{13/2}^+_1)=39$ W.u., and 
$B(E2;{21/2}^+_1\to{17/2}^+_1)=38$ W.u.
Experimentally a $\Delta I=2$ band built on 
the ${15/2}^+$ level at 1333 keV is observed. 
At about the same excitation energy 
a band that starts from the ${11/2}^+_1$ state 
is obtained within the $sdf$-IBFM, 
which is connected by 
the $E2$ transitions of 
$B(E2;{15/2}^+_1\to{11/2}^+_1)=27$ W.u., and 
$B(E2;{19/2}^+_1\to{15/2}^+_1)=30$ W.u.

As in Table~\ref{tab:frac-even}, 
Table~\ref{tab:frac-xe141} gives 
expectation values of the $f$-boson 
$\hat n_f$ [Eq.~(\ref{eq:bh})] 
and single-neutron $\hat n_j$ number operators 
calculated by using the $sdf$-IBFM wave functions. 
It is suggested that 
the low-lying negative-parity states 
are made mainly of the normal-parity 
($\nu 2f_{7/2}$ and $\nu 1h_{9/2}$) configurations, 
while there is practically no contribution 
from $f$ bosons. 
The positive-parity states 
are predominantly made of the 
$\nu 1i_{13/2}$ configuration, i.e., 
$\braket{\hat n_{13/2}}\approx 1$, whereas the 
octupole contributions are negligible, 
except for the ${9/2}^+_2$ state.

\subsubsection{$^{143}$Xe}

As shown in Sec.~\ref{sec:even} 
the RHB-SCMF result suggests a non-zero octupole 
minimum on the $(\beta_2,\beta_3)$-PES for the 
even-even nucleus $^{142}$Xe, and hence the octupole 
correlations are supposed to be most pronounced 
in $^{143}$Xe among the considered odd-mass Xe nuclei. 
The proposed level scheme for $^{143}$Xe is depicted 
in Fig.~\ref{fig:xe143}, and 
the structures of the $sdf$-IBFM wave functions 
of the calculated states are shown in Table~\ref{tab:frac-xe143}. 
The experimental data are taken from 
Ref.~\cite{rzacaurban2011}, suggesting the low-lying 
negative-parity $\Delta I=2$ bands built on 
the ${5/2}^-_1$ and ${7/2}^-_1$ states which 
are lying close in energy to each other. 
The present $sdf$-IBFM reproduces these 
observed bands reasonably well. 
The ${9/2}^-_1$ level is here underestimated, 
probably due to a strong level repulsion 
between the first and second ${9/2}^-$ states. 
The calculation predicts additional bands 
built on the ${9/2}^-_2$ state at $E_x = 0.503$ MeV, 
and on the ${3/2}^-_1$ state lying 
close to the ${5/2}^-_1$ ground state. 
Experimental data from Ref.~\cite{rzacaurban2011} 
suggest two bands based on the states 
with spins ${9/2}$ and ${13/2}$, but  
with parity undetermined.

Table~\ref{tab:frac-xe143} shows that 
there are sizable amounts of the $f$-boson 
components in the negative-parity states, 
i.e., the expectation value 
$\braket{\hat n_f}\approx 0.5$. 
The $f$-boson effects are suggested to be 
even more pronounced in the 
positive-parity states, i.e.,  
$\braket{\hat n_f}\approx 1.5$. 
In the present $sdf$-IBFM calculation, 
the positive-parity bands based on the 
${5/2}^+_1$, ${7/2}^+_1$, and ${9/2}^+_2$ band heads 
are suggested to be of octupole nature 
(highlighted as thick lines in Fig.~\ref{fig:xe143}), 
since in the wave functions of those states belonging 
to these bands significant amounts of the $f$-boson 
components are present, with the 
expectation values typically 
$\braket{\hat n_f}\approx 1.4-1.7$ (see Table~\ref{tab:frac-xe143}).

%
%
\begin{table}[htb!]
\caption{\label{tab:frac-xe143}
Similar to the caption to Table~\ref{tab:frac-xe141}, but for $^{143}$Xe.}
\begin{center}
\begin{ruledtabular}
\begin{tabular}{cccccccc}
\multirow{2}{*}{$I^{\pi}$} &
\multirow{2}{*}{$\braket{\hat n_f}$} &
\multicolumn{6}{c}{$\braket{n_j}$} \\
\cline{3-8}
& &
\textrm{$3p_{1/2}$} &
\textrm{$3p_{3/2}$} &
\textrm{$2f_{5/2}$} &
\textrm{$2f_{7/2}$} &
\textrm{$1h_{9/2}$} &
\textrm{$1i_{13/2}$} \\
\hline
$ {3/2}^{-}_{1}$ & $0.615$ & $0.023$ & $0.123$ & $0.013$ & $0.565$ & $0.276$ & $0.000$ \\ 
$ {5/2}^{-}_{1}$ & $0.527$ & $0.001$ & $0.006$ & $0.056$ & $0.055$ & $0.882$ & $0.000$ \\ 
$ {7/2}^{-}_{1}$ & $0.643$ & $0.000$ & $0.003$ & $0.019$ & $0.060$ & $0.917$ & $0.000$ \\ 
$ {7/2}^{-}_{2}$ & $0.553$ & $0.008$ & $0.060$ & $0.018$ & $0.264$ & $0.650$ & $0.000$ \\ 
$ {9/2}^{-}_{1}$ & $0.790$ & $0.001$ & $0.001$ & $0.032$ & $0.043$ & $0.923$ & $0.000$ \\ 
$ {9/2}^{-}_{2}$ & $0.423$ & $0.001$ & $0.013$ & $0.022$ & $0.138$ & $0.826$ & $0.000$ \\ 
$ {11/2}^{-}_{1}$ & $0.624$ & $0.000$ & $0.004$ & $0.015$ & $0.059$ & $0.921$ & $0.000$ \\ 
$ {11/2}^{-}_{2}$ & $0.663$ & $0.022$ & $0.158$ & $0.013$ & $0.484$ & $0.323$ & $0.000$ \\ 
$ {13/2}^{-}_{1}$ & $0.697$ & $0.004$ & $0.001$ & $0.051$ & $0.037$ & $0.907$ & $0.000$ \\ 
$ {13/2}^{-}_{2}$ & $0.408$ & $0.001$ & $0.011$ & $0.023$ & $0.121$ & $0.844$ & $0.000$ \\ 
$ {15/2}^{-}_{1}$ & $0.512$ & $0.000$ & $0.007$ & $0.012$ & $0.075$ & $0.907$ & $0.000$ \\ 
$ {17/2}^{-}_{1}$ & $0.627$ & $0.009$ & $0.001$ & $0.087$ & $0.031$ & $0.871$ & $0.000$ \\ 
$ {17/2}^{-}_{2}$ & $0.711$ & $0.004$ & $0.006$ & $0.062$ & $0.089$ & $0.839$ & $0.000$ \\ 
$ {19/2}^{-}_{1}$ & $0.728$ & $0.000$ & $0.010$ & $0.010$ & $0.086$ & $0.894$ & $0.000$ \\ 
$ {5/2}^{+}_{1}$ & $1.492$ & $0.004$ & $0.006$ & $0.051$ & $0.036$ & $0.902$ & $0.000$ \\ 
$ {7/2}^{+}_{1}$ & $1.477$ & $0.051$ & $0.037$ & $0.084$ & $0.065$ & $0.763$ & $0.000$ \\ 
$ {9/2}^{+}_{1}$ & $1.445$ & $0.003$ & $0.009$ & $0.046$ & $0.057$ & $0.884$ & $0.000$ \\ 
$ {9/2}^{+}_{2}$ & $1.671$ & $0.060$ & $0.317$ & $0.021$ & $0.468$ & $0.134$ & $0.000$ \\ 
$ {11/2}^{+}_{1}$ & $1.429$ & $0.009$ & $0.010$ & $0.031$ & $0.048$ & $0.902$ & $0.000$ \\ 
$ {13/2}^{+}_{1}$ & $1.550$ & $0.002$ & $0.006$ & $0.029$ & $0.048$ & $0.914$ & $0.000$ \\ 
$ {13/2}^{+}_{2}$ & $1.735$ & $0.015$ & $0.182$ & $0.013$ & $0.717$ & $0.075$ & $0.000$ \\ 
$ {15/2}^{+}_{1}$ & $1.712$ & $0.001$ & $0.001$ & $0.031$ & $0.018$ & $0.949$ & $0.000$ \\ 
$ {17/2}^{+}_{1}$ & $1.401$ & $0.001$ & $0.007$ & $0.016$ & $0.061$ & $0.915$ & $0.000$ \\
$ {17/2}^{+}_{2}$ & $1.499$ & $0.011$ & $0.184$ & $0.017$ & $0.707$ & $0.080$ & $0.000$ \\ 
$ {19/2}^{+}_{1}$ & $1.502$ & $0.001$ & $0.000$ & $0.049$ & $0.011$ & $0.939$ & $0.000$ \\ 
\end{tabular}
\end{ruledtabular}
\end{center}
\end{table}

Figure~\ref{fig:143soft} compares the 
$sdf$-IBFM energy spectra calculated with and without 
taking into account the 
$\beta_3\neq 0$ global minimum in the mapping procedure 
for the even-even core $^{142}$Xe. 
The same boson-fermion strength parameters 
are used for the two different $sdf$-IBFM 
calculations, the only difference being in 
the Hamiltonian parameters for the even-even core. 
It is suggested that 
the negative-parity states 
are not affected by the assumption of the 
zero octupole minimum in the mapping, 
since the negative-parity states are mainly 
comprised of normal-parity single-particle 
configurations coupled to the $sd$-boson space, 
whereas the $f$-boson 
contributions are minor, 
$\braket{\hat n_f}\approx 0.5$ 
[cf. Table~\ref{tab:frac-xe143}]. 
On the other hand, the positive-parity levels 
calculated with the $\beta_3= 0$ minimum 
are higher than those with the $\beta_3 \neq 0$ minimum, 
since the positive-parity states for $^{143}$Xe 
are mostly of octupole nature, i.e., $\braket{\hat n_f}\approx 1.5$, 
and are affected by the presence 
of the $\beta_3 \neq 0$ minimum.

\subsubsection{$^{145}$Xe}

The RHB-SCMF calculation for 
the even-even core nucleus $^{144}$Xe suggests 
a $\beta_3$-soft potential in the SCMF calculation, 
so the octupole correlations are still expected to 
play non-negligible roles in the neighboring $^{145}$Xe. 
Figure~\ref{fig:xe145} exhibits the predicted level 
scheme for $^{145}$Xe, and the wave function 
contents of the relevant states are shown 
in Table~\ref{tab:frac-xe145}. 
Note that experimental information is currently 
not available for $^{145}$Xe. 
Figure~\ref{fig:xe145} displays four negative-parity 
$\Delta I=2$ bands connected by strong $E2$ 
transitions. 
As one sees in Table~\ref{tab:frac-xe145}, 
negative-parity states are mostly composed of the 
$\nu 2f_{7/2}$ and $\nu 1h_{9/2}$ normal-parity 
configurations coupled to the $sd$-boson space, whereas 
contributions from $f$ bosons are present but not 
significant, since 
as the expectation values within the range 
$\braket{\hat n_f}\approx 0.2-0.4$.

%
%
\begin{table}[htb!]
\caption{\label{tab:frac-xe145}
Similar to the caption to Table~\ref{tab:frac-xe141}, but for $^{145}$Xe.}
\begin{center}
\begin{ruledtabular}
\begin{tabular}{cccccccc}
\multirow{2}{*}{$I^{\pi}$} &
\multirow{2}{*}{$\braket{\hat n_f}$} &
\multicolumn{6}{c}{$\braket{n_j}$} \\
\cline{3-8}
& &
\textrm{$3p_{1/2}$} &
\textrm{$3p_{3/2}$} &
\textrm{$2f_{5/2}$} &
\textrm{$2f_{7/2}$} &
\textrm{$1h_{9/2}$} &
\textrm{$1i_{13/2}$} \\
\hline
$ {3/2}^{-}_{1}$ & $0.425$ & $0.004$ & $0.111$ & $0.001$ & $0.836$ & $0.045$ & $0.002$ \\ 
$ {5/2}^{-}_{1}$ & $0.402$ & $0.000$ & $0.019$ & $0.004$ & $0.947$ & $0.028$ & $0.002$ \\ 
$ {7/2}^{-}_{1}$ & $0.491$ & $0.001$ & $0.050$ & $0.002$ & $0.915$ & $0.030$ & $0.002$ \\ 
$ {7/2}^{-}_{2}$ & $0.400$ & $0.000$ & $0.001$ & $0.009$ & $0.152$ & $0.837$ & $0.000$ \\ 
$ {7/2}^{-}_{3}$ & $0.386$ & $0.004$ & $0.088$ & $0.003$ & $0.838$ & $0.066$ & $0.002$ \\ 
$ {9/2}^{-}_{1}$ & $0.445$ & $0.000$ & $0.004$ & $0.008$ & $0.136$ & $0.851$ & $0.000$ \\ 
$ {9/2}^{-}_{2}$ & $0.411$ & $0.000$ & $0.012$ & $0.007$ & $0.861$ & $0.117$ & $0.002$ \\ 
$ {11/2}^{-}_{1}$ & $0.433$ & $0.001$ & $0.048$ & $0.001$ & $0.814$ & $0.134$ & $0.002$ \\ 
$ {11/2}^{-}_{2}$ & $0.410$ & $0.000$ & $0.004$ & $0.011$ & $0.154$ & $0.831$ & $0.000$ \\ 
$ {13/2}^{-}_{1}$ & $0.408$ & $0.000$ & $0.002$ & $0.013$ & $0.053$ & $0.932$ & $0.000$ \\ 
$ {13/2}^{-}_{2}$ & $0.328$ & $0.000$ & $0.012$ & $0.006$ & $0.919$ & $0.059$ & $0.004$ \\ 
$ {15/2}^{-}_{1}$ & $0.341$ & $0.000$ & $0.039$ & $0.000$ & $0.706$ & $0.252$ & $0.003$ \\ 
$ {17/2}^{-}_{1}$ & $0.331$ & $0.000$ & $0.001$ & $0.018$ & $0.034$ & $0.947$ & $0.000$ \\ 
$ {17/2}^{-}_{2}$ & $0.245$ & $0.000$ & $0.009$ & $0.006$ & $0.939$ & $0.039$ & $0.007$ \\ 
$ {19/2}^{-}_{1}$ & $0.268$ & $0.000$ & $0.030$ & $0.000$ & $0.640$ & $0.311$ & $0.018$ \\ 
$ {5/2}^{+}_{1}$ & $1.404$ & $0.023$ & $0.134$ & $0.013$ & $0.760$ & $0.035$ & $0.034$ \\ 
$ {9/2}^{+}_{1}$ & $0.431$ & $0.000$ & $0.000$ & $0.000$ & $0.002$ & $0.000$ & $0.998$ \\ 
$ {9/2}^{+}_{2}$ & $1.381$ & $0.002$ & $0.081$ & $0.002$ & $0.902$ & $0.008$ & $0.005$ \\ 
$ {11/2}^{+}_{1}$ & $0.441$ & $0.000$ & $0.000$ & $0.000$ & $0.008$ & $0.000$ & $0.992$ \\ 
$ {13/2}^{+}_{1}$ & $0.516$ & $0.000$ & $0.000$ & $0.000$ & $0.002$ & $0.000$ & $0.997$ \\ 
$ {13/2}^{+}_{2}$ & $1.423$ & $0.000$ & $0.046$ & $0.002$ & $0.925$ & $0.007$ & $0.020$ \\ 
$ {15/2}^{+}_{1}$ & $0.444$ & $0.000$ & $0.000$ & $0.000$ & $0.006$ & $0.001$ & $0.994$ \\ 
$ {17/2}^{+}_{1}$ & $0.432$ & $0.000$ & $0.000$ & $0.000$ & $0.003$ & $0.000$ & $0.997$ \\ 
$ {17/2}^{+}_{2}$ & $1.318$ & $0.000$ & $0.035$ & $0.001$ & $0.905$ & $0.025$ & $0.033$ \\ 
$ {19/2}^{+}_{1}$ & $0.350$ & $0.000$ & $0.000$ & $0.000$ & $0.006$ & $0.001$ & $0.993$ \\ 
$ {21/2}^{+}_{1}$ & $0.324$ & $0.000$ & $0.000$ & $0.000$ & $0.002$ & $0.000$ & $0.997$ \\ 
$ {23/2}^{+}_{1}$ & $0.255$ & $0.000$ & $0.000$ & $0.000$ & $0.005$ & $0.001$ & $0.993$ \\
\end{tabular}
\end{ruledtabular}
\end{center}
\end{table}

From Table~\ref{tab:frac-xe145}, 
for the positive-parity yrast states of $^{145}$Xe, 
such as the ${9/2}^+_1$, ${11/2}^+_1$ and ${13/2}^+_1$ states, 
the contributions from the $\nu 1i_{13/2}$ 
single-neutron configurations play much more 
important roles in many of the positive-parity states 
than in $^{143}$Xe, while the $f$ boson contents are 
relatively small, 
$\braket{\hat n_f}\approx 0.4-0.5$. 
The $sdf$-IBFM produces the low-lying 
positive-parity ${13/2}^+_1$ and ${11/2}^+_1$ bands, 
with the bandheads being at $E_x \approx 0.493$ MeV and 
0.946 MeV, respectively. These bands are mainly made 
of the $sd$-boson coupled mainly to the $\nu 1h_{9/2}$ 
single-particle orbital (see Table~\ref{tab:frac-xe145}). 
The $f$-boson components make significant contributions 
to determine the wave functions of the non-yrast 
positive-parity states, 
$\braket{\hat n_f}\approx 1.3-1.4$. 
The ${5/2}^+_1$ level at 933 keV 
is primarily of octupole nature, 
$\braket{\hat n_f}\approx 1.4$, similarly to $^{143}$Xe. 
The band built on this state, 
shown in thick lines in Fig.~\ref{fig:xe145}, 
is of octupole nature as it turns out to be 
$\braket{\hat n_f}\approx 1.3-1.4$ for the 
corresponding $sdf$-IBFM wave functions.

\subsection{Transition properties}

\begin{table}[hb!]
\caption{\label{tab:tr-odd}
Predicted $B(E2)$ and $B(E3)$ transition rates 
(in W.u.) for the low-lying states of 
the odd-mass $^{141,143,145}$Xe nuclei, 
calculated with the $sdf$-IBFM.}
\begin{center}
 \begin{ruledtabular}
\begin{tabular}{cccc}
\textrm{$B(E\lambda;I^{\pi}_i \to I^{\pi}_f)$} &
\textrm{$^{141}$Xe} &
\textrm{$^{143}$Xe} &
\textrm{$^{145}$Xe} \\
\hline
$ B(E2; {3/2}^{-}_{1} \to {5/2}^{-}_{1})$ & $1.3$ & $19.7$ & $30.96$ \\ 
$ B(E2; {3/2}^{-}_{1} \to {7/2}^{-}_{1})$ & $0.5$ & $5.9$ & $54.61$ \\ 
$ B(E2; {7/2}^{-}_{1} \to {5/2}^{-}_{1})$ & $0.2$ & $48.5$ & $43.25$ \\ 
$ B(E2; {9/2}^{-}_{1} \to {5/2}^{-}_{1})$ & $0.1$ & $22.8$ & $6.02$ \\ 
$ B(E2; {11/2}^{-}_{1} \to {7/2}^{-}_{1})$ & $6.1$ & $39.2$ & $47.85$ \\ 
$ B(E2; {13/2}^{-}_{1} \to {9/2}^{-}_{1})$ & $7.8$ & $61.9$ & $38.77$ \\ 
$ B(E2; {15/2}^{-}_{1} \to {11/2}^{-}_{1})$ & $2.7$ & $68.9$ & $57.19$ \\ 
$ B(E2; {17/2}^{-}_{1} \to {13/2}^{-}_{1})$ & $0.6$ & $77.2$ & $51.58$ \\ 
$ B(E2; {9/2}^{+}_{1} \to {5/2}^{+}_{1})$ & $23.0$ & $18.2$ & $1.59$ \\ 
$ B(E2; {11/2}^{+}_{1} \to {7/2}^{+}_{1})$ & $5.6$ & $25.1$ & $0.68$ \\ 
$ B(E2; {13/2}^{+}_{1} \to {9/2}^{+}_{1})$ & $34.9$ & $45.2$ & $54.01$ \\ 
$ B(E2; {15/2}^{+}_{1} \to {11/2}^{+}_{1})$ & $26.6$ & $42.9$ & $42.74$ \\ 
$ B(E2; {17/2}^{+}_{1} \to {13/2}^{+}_{1})$ & $39.0$ & $56.9$ & $59.97$ \\ 
$ B(E2; {19/2}^{+}_{1} \to {15/2}^{+}_{1})$ & $30.4$ & $64.5$ & $50.79$ \\ 
$ B(E2; {21/2}^{+}_{1} \to {17/2}^{+}_{1})$ & $37.7$ & $61.6$ & $61.78$ \\
$ B(E3; {9/2}^{+}_{1} \to {3/2}^{-}_{1})$ & $2.9$ & $2.2$ & $5.54$ \\ 
$ B(E3; {11/2}^{+}_{1} \to {5/2}^{-}_{1})$ & $0.2$ & $14.0$ & $4.64$ \\ 
$ B(E3; {13/2}^{+}_{1} \to {7/2}^{-}_{1})$ & $0.0$ & $18.5$ & $8.95$ \\ 
$ B(E3; {15/2}^{+}_{1} \to {9/2}^{-}_{1})$ & $0.0$ & $23.3$ & $0.78$ \\ 
$ B(E3; {17/2}^{+}_{1} \to {11/2}^{-}_{1})$ & $0.0$ & $33.3$ & $9.23$ \\ 
\end{tabular}
 \end{ruledtabular}
\end{center}
\end{table}

Table~\ref{tab:tr-odd} shows  
the calculated $B(E2)$ and $B(E3)$ 
transition strengths for yrast states of the 
odd-mass $^{141,143,145}$Xe nuclei. 
The calculated $B(E2)$ rates for $^{143,145}$Xe are 
generally larger than those for $^{141}$Xe, 
reflecting that the quadrupole collectivity 
increases from $^{141}$Xe to $^{143,145}$Xe isotopes. 
Some transition rates, such as 
the $B(E2; {9/2}^-_1 \to {5/2}^-_1)$, and 
$B(E2; {11/2}^+_1 \to {7/2}^+_1)$ ones 
are predicted to be large in $^{143}$Xe, 
while they are much more smaller in $^{141,145}$Xe, 
suggesting that the nuclear structure of $^{143}$Xe 
differs from the adjacent nuclei $^{141,145}$Xe. 
For all the three nuclei, large $\Delta I=2$ 
in-band $E2$ transition rates for the positive-parity states, 
$B(E2; {13/2}^+_1 \to {9/2}^+_1)$, 
$B(E2; {15/2}^+_1 \to {11/2}^+_1)$,
$B(E2; {17/2}^+_1 \to {13/2}^+_1)$, and  
$B(E2; {19/2}^+_1 \to {15/2}^+_1)$, 
are obtained. 
The $B(E3)$ rates predicted for $^{143}$Xe are 
generally large, as compared to $^{141,145}$Xe, 
which are also considered to be a signature of 
octupole correlations in $^{143}$Xe.

\section{Conclusions\label{sec:summary}}

To summarize, signatures of octupole correlations 
in the neutron-rich odd-mass 
Xe isotopes near $N=88$ have been studied 
within the theoretical framework of the 
nuclear EDF and $sdf$-IBFM. 
The $sdf$-IBM Hamiltonian describing quadrupole 
and octupole collective states of the even-even nuclei, 
and single-particle energies 
and occupation probabilities of an unpaired neutron, 
were determined by using the results of the 
RHB-SCMF calculations based on the relativistic EDF. 
Strength parameters for the 
boson-fermion interactions were, however, 
empirically determined so as to reproduce 
the observed low-lying levels in each odd-mass Xe nucleus, 
and energy-level systematic 
in the neighboring odd-mass Ba nuclei.

The microscopic RHB-SCMF calculations suggested 
for $^{142}$Xe 
a potential that is notably soft along the 
octupole $\beta_3$ deformation with 
a shallow, but non-zero octupole global mean-field 
minimum at $\beta_3 \approx 0.05$. 
There is no octupole deformed minimum in $^{140}$Xe 
and $^{144}$Xe, while in the latter nucleus 
the potential was shown to be quite 
soft in the $\beta_3$ direction and 
hence the octupole degree of freedom should not 
be negligible. 
The calculated energy spectra for the even-even 
Xe nuclei showed certain traces of the 
octupole correlations, including the low-energy 
negative-parity states of octupole nature 
at $^{142}$Xe. 
The octupole degree of freedom has been shown to 
have influences on the low-lying spectra for 
the neighboring odd-mass Xe nuclei. 
The evolution of the low-energy spectra as 
functions of $N$ and their wave function contents 
suggested a variation of nuclear structure 
in the vicinity of $N=88$, at which octupole 
collectivity is expected to be most pronounced. 
In particular, many of the low-lying 
positive-parity states 
for $^{143}$Xe were shown to contain 
more than one $f$ bosons in their wave functions, 
and to exhibit sizable amounts of the $E3$ 
transitions to the negative-parity 
ground-state band.

The present work has suggested that the octupole 
correlations still play a part in describing 
those states of the odd-mass Xe isotopes with 
the excitation energies $E_x \gtrsim 0.5$ MeV, 
similarly to the neighboring odd-mass Ba isotopes. 
The results of the present spectroscopic calculations 
would stimulate experimental studies aimed 
to access those neutron-rich Xe isotopes that are 
far from the $N=82$ shell closure, which also have been 
beyond reach of microscopic nuclear structure calculations. 
Possible future studies include an application 
of the $sdf$-IBFM to more challenging cases 
of light actinides, for which static 
octupole deformation is suggested. 
On the other hand, since the boson-fermion strength 
parameters were here obtained using only limited 
experimental information and based on the assumption that 
the low-lying structures of the odd-mass Xe nuclei should 
be similar to those of the neighboring Ba nuclei, 
the reliability of these parameters remains an open question. 
It can be addressed, ideally, by developing a method to 
determine all of the boson-fermion interaction strengths 
in a fully microscopic way, that is, only by using 
results of the EDF-SCMF calculation and without 
referring to the experimental data. 
The works along these lines are in 
progress and will be reported elsewhere.

\acknowledgements
The author thanks Atsuko Odahara and Kenichi Yoshida 
for valuable discussions. 
This work has been supported in part  
by the RCNP Collaboration Research Network 
program as the Project No. COREnet-051.

\bibliography{refs}

\end{document}